\documentclass[final,3p,times,twocolumn]{elsarticle}

\usepackage{amssymb}
\usepackage{amsmath}
\usepackage{lineno}
\usepackage{xspace}
\usepackage{todonotes}
\usepackage{xcolor}
\usepackage{amsthm}
\usepackage{subcaption}
\usepackage{graphicx}
\usepackage{multirow}
\usepackage{booktabs}
\usepackage{url}
\usepackage{hyperref}
\usepackage{fontawesome5}
\usepackage{pifont} 
\usepackage{makecell}
\usepackage[draft]{changes}
\definechangesauthor[name={Silvia}, color=red]{S}
\setauthormarkup{}

\newcommand{\Octree}{Octree\xspace}
\newcommand{\Octrees}{Octrees\xspace}
\newcommand{\KDTree}{KD-tree\xspace}
\newcommand{\KDTrees}{KD-trees\xspace}

\newcommand{\unibnOctreeLib}{\textit{unibnOctree}}
\newcommand{\nanoflann}{\textit{nanoflann}}
\newcommand{\picotree}{\textit{picotree}\xspace}

\newcommand{\ptrOctree}{\textit{ptrOctree}}
\newcommand{\linOctree}{\textit{linOctree}}
\newcommand{\unibnOctree}{\textit{uniOctree}}
\newcommand{\pclOctree}{\textit{pclOctree}}
\newcommand{\pclKDTree}{\textit{pclKD}}
\newcommand{\nanoflannKDTree}{\textit{nanoKD}}
\newcommand{\picotreeKDTree}{\textit{picoKD}\xspace}

\newcommand{\neighboursPtr}{\textit{neighboursPtr}\xspace}
\newcommand{\neighboursLin}{\textit{neighboursLin}\xspace}
\newcommand{\neighboursPrune}{\textit{neighboursPrune}\xspace}
\newcommand{\neighboursStruct}{\textit{neighboursStruct}\xspace}
\newcommand{\neighboursUnibn}{\textit{neighboursUnibn}\xspace}
\newcommand{\neighboursPCLOct}{\textit{neighboursPCLOct}\xspace}
\newcommand{\neighboursPCLKD}{\textit{neighboursPCLKD}\xspace}
\newcommand{\neighboursNano}{\textit{neighboursNano}\xspace}
\newcommand{\neighboursPico}{\textit{neighboursPico}\xspace}

\newcommand{\knnLinOct}{\textit{knnLinOct}\xspace}
\newcommand{\knnPCLOct}{\textit{knnPCLOct}\xspace}
\newcommand{\knnPCLKD}{\textit{knnPCLKD}\xspace}
\newcommand{\knnNano}{\textit{knnNano}\xspace}
\newcommand{\knnPico}{\textit{knnPico}\xspace}

\newcommand{\ptrNone}{\textit{ptrNone}\xspace}
\newcommand{\ptrMort}{\textit{ptrMort}\xspace}
\newcommand{\ptrHilb}{\textit{ptrHilb}\xspace}
\newcommand{\linNone}{\textit{linNone}\xspace}
\newcommand{\linMort}{\textit{linMort}\xspace}
\newcommand{\linHilb}{\textit{linHilb}\xspace}

\newcommand{\kernel}[1]{\mathcal{N}_{\text{#1}}}

\definecolor{octUnencoded}{HTML}{F0E442}  % Okabe-Ito yellow
\definecolor{octMort}{HTML}{E69F00}       % Okabe-Ito orange
\definecolor{octHilb}{HTML}{D55E00}       % Okabe-Ito vermilion
\definecolor{loctMort}{HTML}{56B4E9}      % Okabe-Ito sky blue
\definecolor{loctHilb}{HTML}{0072B2}      % Okabe-Ito blue
\definecolor{neighboursPtr}{HTML}{117733}     % Tol Muted green
\definecolor{neighbours}{HTML}{44AA99}        % Tol Muted teal
\definecolor{neighboursPrune}{HTML}{88CCEE}   % Tol Muted cyan
\definecolor{neighboursStruct}{HTML}{332288}  % Tol Muted indigo
\definecolor{neighboursUnibn}{HTML}{CC6677}   % Tol Muted rose
\definecolor{neighboursPCLOct}{HTML}{888888}  % Tol Muted grey
\definecolor{neighboursPCLKD}{HTML}{AA4499}   % Tol Muted purple
\definecolor{neighboursNano}{HTML}{DDCC77}    % Tol Muted sand
\definecolor{neighboursPico}{HTML}{882255}    % Tol Muted wine

\newcommand{\dotneighboursPtr}{\textcolor{neighboursPtr}{\textbullet}~}
\newcommand{\dotneighbours}{\textcolor{neighbours}{\textbullet}~}
\newcommand{\dotneighboursPrune}{\textcolor{neighboursPrune}{\textbullet}~}
\newcommand{\dotneighboursStruct}{\textcolor{neighboursStruct}{\textbullet}~}

\newtheorem{theorem}{Theorem}
\theoremstyle{definition}
\newtheorem{definition}[theorem]{Definition}

\journal{Information Sciences}

\begin{document}

\begin{frontmatter}



\title{Efficient Neighbourhood Search in 3D Point Clouds Through Space-Filling Curves and Linear Octrees}


\author[citius]{Pablo D. Viñambres} %% Author name
\author[citius,dec]{Miguel Yermo\corref{cor}}
\author[eth]{Silvia R. Alcaraz}
\author[citius,dec]{Oscar G. Lorenzo}
\author[citius,dec]{Francisco F. Rivera}
\author[citius,dec]{José C. Cabaleiro}

\cortext[cor]{Corresponding author: miguel.yermo@usc.es}

\affiliation[citius]{organization={Centro de Investigación en Tecnoloxías Intelixentes (CiTIUS)},%Department and Organization
            addressline={Rúa de Jenaro de la Fuente Domínguez}, 
            city={Santiago de Compostela},
            postcode={15782}, 
            state={Galicia},
            country={Spain}
}
\affiliation[dec]{organization={Departamento de Electrónica e Computación, Escola Técnica Superior de Enxeñaría, Universidade de Santiago de Compostela},%Department and Organization
            addressline={Rúa Lope Gómez de Marzoa}, 
            city={Santiago de Compostela},
            postcode={15782}, 
            state={Galicia},
            country={Spain}
}
\affiliation[eth]{organization={Systems Group, Department of Computer Science, ETH Zurich},
            addressline={Stampfenbachstrasse 114},
            city={Zürich},
            postcode={8092},
            country={Switzerland}
}

%Systems Group, Department of Computer Science, ETH Zurich, 8092 Zurich, Switzerland

%% Abstract
\begin{abstract}

%% Limit (Fut Gen Com Sys): 250 words. Now: 210.
This work presents an efficient approach for neighbourhood searching in 3D point clouds by combining spatial reordering leveraging Space-Filling Curves (SFCs), specifically Morton and Hilbert curves, with a linear \Octree implementation. 
We also propose specialised search algorithms for fixed-radius and kNN queries, based on our linear \Octree structures. 
Additionally, we introduce the kNN locality histogram as a diagnostic metric for data access locality, establishing that its skewness directly correlates with cache misses and search performance.
Our experiments reveal that SFC reordering significantly improves spatial data access, reducing cache misses from 25\% to 75\% and runtime by up to 50\%. 
We evaluate our proposal against several widely used \Octree and \KDTree implementations across diverse, large-scale public LiDAR datasets.
Our most efficient structure (\neighboursStruct) achieves up to 10$\times$ faster search times compared to existing solutions for large-radius queries; this gain combines algorithmic enhancements with a range-based output format that avoids coordinate materialisation.
Furthermore, our OpenMP-parallelised searches yield high scalability across cores and problem sizes, demonstrating a speedup of up to $36\times$ using a fixed-radius on a 40-core system.
The results indicate that our methods provide a robust and efficient solution for applications requiring fast access to large-scale 3D point neighbour sets.
\end{abstract}

%% Keywords
\begin{keyword}
%% Future Generations:  6 to 10 keywords
%% Information Science: You are required to provide 1 to 7 keywords for indexing purposes. Keywords should be written in English. Please try to avoid keywords consisting of multiple words (using "and" or "of").
Octree \sep Space-Filling Curves \sep Spatial Partitioning \sep Neighbourhood Searching \sep 3D LiDAR Point Clouds \sep Data Locality \sep Parallel processing


\end{keyword}

\end{frontmatter}

%\linenumbers

\section{Introduction} % 1

Light Detection and Ranging (LiDAR) and photogrammetry technologies have become the standard techniques for acquiring 3D point clouds across diverse applications, including mapping, urban modelling, autonomous driving, and archaeology. 
Continuous advances in these acquisition techniques have led to a notable increase in the volume and density of the data acquired. 
Consequently, the efficient extraction of meaningful information from such large-scale point clouds has emerged as a significant computational challenge.

In this context, much of the relevant information is encoded in the local geometric relationships between the points of the cloud.
Fundamental tasks such as clustering, segmentation, feature extraction, and data mining often rely on finding the neighbourhood of a given point. Beyond the computational cost arising from performing neighbourhood queries over large-scale 3D point clouds, the inherently unstructured and irregular nature of such data further amplifies the complexity.
Thus, the design of efficient neighbourhood search algorithms becomes essential.

Traditionally, neighbour searching in 3D point clouds has been addressed using hierarchical data structures such as \KDTrees and \Octrees, which limit computations to regions near the query point. 
Nevertheless, the performance associated with these structures can be affected by the spatial and memory distribution of the data. 
Since neighbourhood computation involves identifying spatially close points, it is desirable to ensure that these points are also close in memory. 
Spatial reordering techniques based on Space-Filling Curves (SFCs), such as the Morton and Hilbert curves, map three-dimensional points to a one-dimensional representation while preserving spatial locality. 

In this work, we demonstrate that SFC-based reordering techniques improve memory access coherence and enhance the performance of neighbour searches when using \Octrees and \KDTrees.
Furthermore, we present an efficient method for neighbourhood search in 3D point clouds which combines spatial reorderings based on Morton and Hilbert curves with a linear \Octree implementation.
In our experiments, we compare different data structures, including pointer-based \Octrees, \KDTrees, and our linear \Octree.
For the latter, we assess an improved fixed-radius neighbourhood search algorithm that leverages and expands the structure proposed by Keller et al.~\cite{Keller2023}.

While the linear \Octree adopted in this work builds on the compact array-based structure introduced by Keller et al.~\cite{Keller2023} for particle simulations in distributed multi-node environments, our contributions are distinct in domain and algorithmic scope. Keller et al.'s structure was designed for uniformly distributed synthetic data with exclusively Hilbert curve ordering and no neighbourhood search optimisations. In contrast, this work targets real-world LiDAR point clouds, which present structural challenges absent in particle simulations: non-uniform density gradients, scan-pattern artefacts, and wide spatial extent. Beyond adapting the data structure to this domain, our specific algorithmic contributions are: (i) the \textit{internalRanges} extension, which enables direct index-range access to any octant; (ii) the \neighboursPrune algorithm, which exploits full-containment early termination for fixed-radius queries; (iii) the \neighboursStruct algorithm, which returns contiguous index ranges instead of point coordinates, reducing output allocation costs; (iv) a kNN depth-first search adapted to the linear \Octree and its evaluation against other solutions in the literature; (v) support for four search kernel geometries; and (vi) the kNN locality histogram, a diagnostic metric that quantifies the effect of SFC reordering on data locality and links it to measurable cache miss reductions.

Additionally, we parallelise the neighbourhood searches for shared-memory systems using the OpenMP~\cite{openmp_website} library and examine the parallel efficiency and scalability of each method for both random and sequential queries across a set of heterogeneous, openly available point cloud datasets.

The rest of this manuscript is organised as follows: Section~\ref{sec:state-of-the-art} reviews prior work on neighbourhood queries in 3D point clouds, comparing our proposal with existing approaches. Section~\ref{sec:computing-neighborhoods-in-3D-pointcloud} explains the neighbour search procedure according to the underlying data structure; Section~\ref{sec:space-filling-curves} introduces SFCs and, in particular, Morton and Hilbert orderings; Section~\ref{sec:enhancing-performance-through-data-locality} describes our method to improve query efficiency and defines a diagnostic metric to quantify data locality; Section~\ref{sec:exp_env} outlines the experimental configuration; Section~\ref{sec:results} presents and analyses the results obtained; and, finally, Section~\ref{sec:conclusions} summarises the main conclusions.

\section{Related work}\label{sec:state-of-the-art} % 2

\begin{table*}[htb]
\centering
\caption{Comparison of our proposal against the most relevant state-of-the-art works. Table footnotes clarify cases marked as Partial or Limited.}\label{tab:related-work}
\footnotesize
{
\setlength{\tabcolsep}{3pt}
\begin{tabular}{lllllllll}
\toprule
 & \textbf{This work} & \textbf{Behley~\cite{behley2015efficient}} & \textbf{Keller~\cite{Keller2023}} & \textbf{Guan~\cite{Guan2018}} & \textbf{Wang~\cite{Wang2021}} & \textbf{PCL~\cite{PCL2011}} & \textbf{nanoflann~\cite{Nanoflann2014}} & \textbf{picotree~\cite{picotree}} \\
\midrule
Target domain        & LiDAR      & LiDAR       & Particles     & Database  & LiDAR          & General    & General   & General  \\
SFC ordering         & Mort.+Hilb.& Morton      & Mort.+Hilb.   & Mort.+Hilb.   & Morton    & No   & No  & No \\
Data structure       & Lin. Oct.  & Ptr. Oct.   & Lin. Oct.     & -    & Oct.+R*-Tree   & Ptr. Oct.  & KD-Tree   & KD-Tree  \\
Fixed-radius search kernels & \checkmark & \checkmark & \ding{55} & \ding{55} & \ding{55} & \checkmark & \checkmark & \checkmark \\
kNN search           & \checkmark & \ding{55}    & \ding{55}      & \ding{55}  & \checkmark     & \checkmark & \checkmark & \checkmark \\
Parallelism          & OpenMP     & \ding{55}    & MPI+GPU       & Partial$^{a}$         & \ding{55}       & Partial$^{b}$          & \ding{55}  & \ding{55} \\
Cache miss analysis  & \checkmark & \ding{55}    & \ding{55}      & \ding{55}  & \ding{55}       & \ding{55}   & \ding{55}  & \ding{55} \\
Memory footprint analysis & \checkmark & \ding{55} & \ding{55}   & \ding{55}  & \ding{55}       & \ding{55}   & \ding{55}  & \ding{55} \\
Parallel scalability analysis & \checkmark & \ding{55} & Partial$^{c}$ & Partial$^{d}$  & \ding{55}       & \ding{55}   & \ding{55}  & \ding{55} \\
Multi-dataset evaluation & \checkmark & Limited$^{e}$ & \ding{55} & Limited$^{f}$ & \ding{55}  & \ding{55}   & \ding{55}  & \ding{55} \\
Open-source code     & \checkmark & \checkmark  & \checkmark    & \checkmark  & \ding{55}       & \checkmark & \checkmark & \checkmark \\
\bottomrule
\end{tabular}
}
{
\flushleft 
\footnotesize 
$^{a}$ Parallelization applied to SFC encoding and bulk loading, but not to spatial queries.\\
$^{b}$ OpenMP/TBB support at the algorithm level, not specifically optimised for index traversal.\\
$^{c}$ Evaluates scalability solely for distributed data structure construction.\\
$^{d}$ Reports data loading speedup but lacks query scalability evaluation.\\
$^{e}$ Evaluated using a small set of LiDAR scans.\\
$^{f}$ Restricted to a single-database deployment context.\\ }
\end{table*}

The use of \Octrees and SFCs, such as Morton and Hilbert curves, for managing and analysing 3D point clouds has grown significantly in recent years, driven by the need to improve efficiency in handling large volumes of spatial data. 
\Octrees, which recursively divide 3D space into eight octants, benefit from these curves by mapping multidimensional data into a one-dimensional order, optimising applications such as compression, indexing, and spatial queries, including nearest neighbour searches. 
Both curves present complementary strengths, and their adoption depends on the specific requirements of each application~\cite{Tero2012, Yuening2025}.

Morton codes, known for their simplicity and speed, are widely used in \Octree structures for indexing and compression tasks. Their integration into linear \Octrees has been key to improving efficiency, as observed in the work by Behley et al.~\cite{behley2015efficient}, where they are used to sort points and reduce the need to traverse the tree to its full depth during neighbourhood searches, outperforming popular libraries. In the field of compression, Cao et al.~\cite{Cao2025} propose a method based on autoencoders that transforms point clouds into Morton codes using a linear \Octree, inspired by their effectiveness in visual analysis and image compression tasks. Furthermore, Yang et al.~\cite{Yang2024} highlight how Morton codes facilitate adaptive partitioning strategies, ensuring data locality and improving performance in non-relational databases. In popular software applications, such as CloudCompare~\cite{CloudCompare2023}, Morton-type encodings are used for fast binary searches and spatial sweeps, while Wang et al.~\cite{Wang2021} apply them in a hybrid \Octree and 3D R*-Tree structure to optimise nearest neighbour queries.

Hilbert curves have gained increasing popularity due to their superior performance in preserving spatial locality when indexing three-dimensional point clouds compared to Morton codes. In particular, some studies have demonstrated the superiority of Hilbert curves over Morton curves in maintaining locality during point cloud analysis~\cite{Shao2024}. Recent research highlights the applicability of Hilbert curves for improving data compression efficiency~\cite{Pavlovic2019, Chen2022}. In~\cite{Chen2022b}, the use of Hilbert curves is proposed to reduce the dimensionality of a 3D point cloud, allowing operations to be applied directly in two dimensions, which would otherwise be too costly.

Notwithstanding, to the best of our knowledge, the application of Morton codes or Hilbert curves for neighbourhood searching in 3D point clouds obtained with LiDAR sensors has not been extensively studied. In~\cite{behley2015efficient}, the use of Morton codes is proposed to order points in such a way that it is not necessary to traverse the \Octree to its full depth when certain conditions are met. The results demonstrate better performance than the most popular libraries for neighbour searching. Guan et al.~\cite{Guan2018} present a C++ library for handling massive point clouds in parallel using SFCs. However, the library is oriented towards use with relational databases. Regarding spatial queries, the authors demonstrate that SFCs, particularly the Hilbert curve, outperform classic spatial query methods (such as the B-Tree) when managing a few hundred million data points. Furthermore, Keller et al.~\cite{Keller2023} present an \Octree construction method based on linear arrays built from Hilbert curves for use in particle simulations in distributed environments.

Despite the advantages offered by spatial reordering, many general-purpose point cloud processing workflows do not inherently incorporate memory-locality-preserving reordering. To determine a baseline for comparison, we note that several high-performance libraries have established themselves as benchmarks for spatial queries. The Point Cloud Library (PCL)~\cite{PCL2011} remains one of the most widely used frameworks, providing robust implementations of both pointer-based \Octrees and \KDTrees. In the domain of nearest neighbour searches, the evolution from the versatile FLANN library~\cite{Flann2009} to its header-only fork nanoflann~\cite{Nanoflann2014} has set high standards for query efficiency. More recently, developments such as PicoTree~\cite{picotree} have pushed these boundaries even further, optimising build and search times through modern C++ features. These libraries constitute the current state of the art in spatial indexing, serving as the primary references for assessing the performance gains introduced by the SFC-based methods discussed in this work.

Table~\ref{tab:related-work} summarises the key characteristics of the most relevant prior work and baseline libraries, highlighting the unique combination of features offered by this work.

A related line of research has explored GPU-accelerated spatial search. 
Libraries such as FAISS~\cite{johnson2019billion} target high-throughput approximate nearest-neighbour search on dense vector data, while cuSpatial~\cite{cuspatial:25.04} provides GPU-native spatial join and proximity operations for geospatial workloads.
GPU and multi-GPU accelerated approaches for kNN in 3D point clouds have also been proposed, targeting applications such as point cloud registration, smoothing, and normal estimation~\cite{chang2025accelerating,agathos2024multigpu}. 
%However, these approaches are constrained by GPU memory capacity: large LiDAR clouds such as those used in this work (up to 721 million points, approximately 23~GB) may not fit in a single GPU, requiring either multi-GPU setups or chunked transfers that add latency. 
These approaches may deliver high throughput by leveraging the massive parallelism of GPU architectures, provided that communications do not introduce a bottleneck. 
Otherwise, this overhead could represent a major limitation for latency-sensitive or memory-constrained applications.
%The CPU-based approach presented in this work avoids this constraint, targeting shared-memory HPC systems where the cache-locality gains of SFC reordering directly reduce main-memory bandwidth pressure.}

In contrast, the CPU-based approach presented in this work targets shared-memory HPC systems, providing an efficient solution for CPU-centric nodes or heterogeneous environments where GPU resources are constrained or unavailable. In these scenarios, the cache-locality gains achieved through our SFC reordering directly reduce main-memory bandwidth pressure. 

\section{Computing neighbourhoods in 3D point clouds}\label{sec:computing-neighborhoods-in-3D-pointcloud} % 3
Efficient neighbour searching in massive point clouds relies heavily on spatial decomposition data structures. This study evaluates spatial queries using two Octree-based approaches. First, we establish a baseline with a pointer-based Octree, following the traditional hierarchical branch-and-leaf architecture. Second, we introduce an optimised linear \Octree that leverages reordered point data and a linear implementation of the Octree. Thus, spatial locality and memory throughput is maximized, significantly reducing the computational overhead of neighbour queries.

\subsection{Pointer-based Octree}\label{sec:ptr_octree} % 3.1
\begin{figure}[htb]
    \centering
    \includegraphics[width=0.98\linewidth]{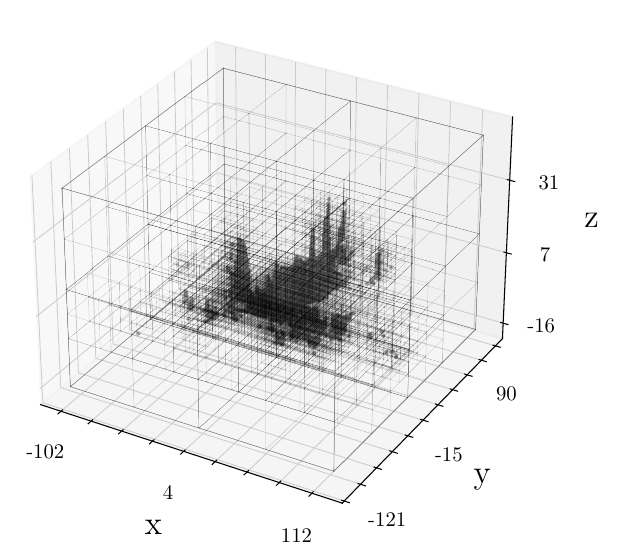}
    \caption{Edges of the octants of an \Octree over point cloud dataset \texttt{bildstein\_station1}~\cite{hackel2017isprs}, up to depth 5. Distances are shown in metres.}
    \label{fig:octree-bildstein-station1}
\end{figure}

\Octrees are tree-like data structures in which each node has exactly eight children. A node with no children is considered a leaf and must contain data. \Octrees are useful for partitioning space: their root contains the entire volume of the bounding box enclosing a 3D point cloud. Each node is iteratively partitioned into eight nodes by halving the length of each of its sides. This process is repeated until the number of points per node does not exceed a threshold, $N_{max}$. In some implementations, an internal leaf is allowed to have fewer than 8 children when the corresponding subspaces contain no points, whereas in others, all 8 children are always created after exceeding $N_{max}$. The former one is called \textit{non-complete} \Octrees, and the latter \textit{complete} \Octrees. An example of the edges of each of these octants is shown in Figure~\ref{fig:octree-bildstein-station1}, which are further subdivided in the denser areas of the point cloud.

Note that when performing a spatial query in this data structure, such as searching for the neighbourhood of a given point, the levels of an \Octree must be traversed sequentially, searching for the branches that contain the leaves of interest. 

\subsection{Fixed-radius queries} % 3.2
Throughout this study, \textit{search kernels} are used to express these spatial queries in 3D point clouds.

\begin{definition}
A \textit{search kernel} is a map $\mathcal{N}: (P, \mathbb{R}^3, \mathbb{R}) \to \mathcal{P}(P)$ that obtains the neighbourhood of a point, considering a specific geometry, from the point cloud $P$, a centre $c \in \mathbb{R}^3$ and a radius $r \in \mathbb{R}$. The following four kernels will be considered in this article:
\begin{itemize}
    \item \textbf{Sphere}: $\kernel{Sphere}(P, c, r) = \{p \in P : \|(p_x-c_x, p_y-c_y, p_z-c_z)\|_2 < r\}.$
    \item \textbf{Circle}: $\kernel{Circle}(P, c, r) = \{p \in P : \|(p_x-c_x, p_y-c_y, 0)\|_2 < r\}.$
    \item \textbf{Cube}: $\kernel{Cube}(P, c, r) = \{p \in P : \|(p_x-c_x, p_y-c_y, p_z-c_z)\|_\infty < r\}.$
    \item \textbf{Square}: $\kernel{Square}(P, c, r) = \{p \in P : \|(p_x-c_x, p_y-c_y, 0)\|_\infty < r\}.$
\end{itemize}

\end{definition}
 
Given a search kernel, an initial straightforward algorithm can be written for searches in any type of \Octree:
\begin{enumerate}
\item If the current node is a branch, obtain which child nodes intersect with that branch.
\item For each intersecting child node, repeat the process until we reach the leaves.
\item For each intersecting leaf, check point by point whether it belongs to the proposed kernel or not.
\item Return all the points within the kernel as an array of point cloud indices or pointers.
\end{enumerate}

On dense point clouds, \Octrees can get deep, making the access operations significantly more expensive. This is because the data structure stores the memory addresses of the corresponding points in each leaf as pointers. Since point clouds are unstructured, points that are close in space may be located in distant parts of memory, leading to numerous cache misses. Moreover, it is not possible to directly recover all points in a branch; therefore, even when the kernel fully contains a branch, checks over all its children must still be performed.

Though there are some ways to circumvent those problems in a pointer-based tree, such as the techniques used in \cite{behley2015efficient}, we will address them more effectively by switching to a new implementation of the \Octree.

\subsection{kNN queries} % 3.3

Though \KDTrees are most frequently used for kNN searches, especially in high-dimensional spaces, \Octrees may also be adapted for this goal in 3D point clouds, which is the approach we use for the linear \Octree introduced in Section~\ref{sec:enhancing-performance-through-data-locality}. We follow the usual approach, keeping a priority queue $Q$ that can contain both single points or octants. Elements of $Q$ are tuples $(i, \delta, o, d)$ where $i$ is the octant or point index, $\delta$ is the depth of the octant (or $\delta = 0$ if the element is a point), $o$ is a boolean indicating whether the element is an octant or a point and $d$ is the squared distance from the element to the search centre.

Our implementation of kNN searches %, which we will call \textit{linOctKNN}, 
extracts the top element $t$ with the lowest $d$ (in constant time) until we have found all $k$ neighbours or the queue is empty, and performs one of the following actions:
\begin{itemize}
    \item If $t$ is a point, insert it into the result list.
    \item If $t$ is a leaf node, insert all its points into $Q$.
    \item If $t$ is an internal node, insert all its child octants into $Q$. 
\end{itemize}

Given an octant with bounds $(x_{m}, x_{M}) \times (y_{m}, y_{M}) \times (z_{m}, z_{M})$ and a point $p = (x, y, z)$, computing the octant-to-point distance is simple due to bounding box alignment. We first find the closest point inside the octant by clamping it to its bounds
\[
\gamma = \begin{pmatrix}
  \min(\max(x, x_m), x_M) \\
  \min(\max(y, y_m), y_M) \\
  \min(\max(z, z_m), z_M)
\end{pmatrix}.
\]

Then, the distance is given by $d = \left\lVert p-\gamma \right\rVert_2$. Note that if $p$ lies inside the octant, then $\gamma = p$ and so $d = 0$. Otherwise, $\gamma$ lies on the closest face, edge, or vertex of the octant.

\section{Space-Filling Curves}\label{sec:space-filling-curves} % 4

In this section, we will explain our implementation and use of the two most common Space-Filling Curves (SFCs). While efficient data structures are essential to the fast retrieval of neighbourhood sets in point clouds, the order in which the points are stored should also be considered. Proper ordering of points can significantly improve the performance of neighbourhood searches by enhancing spatial locality, the property that points in three-dimensional space that are close to each other are also stored closely in memory. To achieve this, we define a map from the three-dimensional point cloud space to the one-dimensional memory space in which it is stored as in~\cite{asano1997space}.

Consider a point cloud $P \subset {\mathbb{R}}^3$, with lower corner $p_{m} = (x_m, y_m, z_m)$ and upper corner $p_{M} = (x_M, y_M, z_M)$, satisfying $x_m \leq x \leq x_M, y_m \leq y \leq y_M, z_m \leq z \leq z_M$ for each point $(x,y,z) \in P$. We define $B = [x_m, x_M) \times [y_m, y_M) \times [z_m, z_M)$. We set a level of subdivision $L \in \mathbb{N}$ such that $N = 2^L$, and consider the grid $S_L = [0, N) \times [0, N) \times [0, N) \subset  \mathbb{N}^3$, $|S_L| = 2^{3L}$. We seek to discretise the bounding box $B$ so that each point $p \in P$ corresponds to a cell $e \in S_L$. To do this, we take the following steps:

\begin{enumerate}
    \item Translate and scale the coordinates of each of the points in the cloud to the cube $[0, N) \times [0, N) \times [0, N) \subset \mathbb{R}^3$.
    \item Convert each of the transformed points into elements of $S_L$ using the floor function. In this way, the coordinates $(\texttt{x}, \texttt{y}, \texttt{z})$ are represented by $L$ bit integers, relating to the cloud bounding box. For instance, the lower corner of $B$ becomes 0, that is:
    $$
        (x_m, y_m, z_m) \to  (\underbrace{\texttt{000} \dots \texttt{0}}_{L \text{ bits}}, \underbrace{\texttt{000} \dots \texttt{0}}_{L \text{ bits}}, \underbrace{\texttt{000} \dots \texttt{0}}_{L \text{ bits}}).
    $$
\end{enumerate}

Since $|S_L| = 2^{3L}$, each of the cells $e$ can be represented by a $3L$ bit code. Finding a way to assign each of these codes to each cell to maximise spatial locality is the main issue covered in this section.

In practice, we use $L=21$ to store each code in a 64-bit integer, discarding the first bit.
%%The discretised coordinates $(\texttt{x}, \texttt{y}, \texttt{z})$ are then stored in 32-bit integers,  ignoring the first 11 bits. 
With this precision, it is unlikely that more than one point will be assigned to the same cell, although this may occur without causing any problems in practice. Thus, the method for assigning codes to cells is based on SFCs, which are defined below.

\begin{definition}
    A Space-Filling Curve (SFC) of level $l$ is a mapping $C: S_l \to [0, 2^{3l})$. An SFC of level $l$ is said to be \textit{recursive} if $l=0$ or if $l>0$ and we can partition $C$ into eight recursive SFCs $C_0, \dots C_7$  of level $l-1$ such that there exists a permutation $\pi: \{0, \dots, 7\} \to \{0, \dots, 7\}$ for which $C_{\pi(i)}$ contains the $i$-th range of codes at level $l-1$, $\{i2^{3(l-1)}, \dots, (i+1)2^{3(l-1)} -1\}$, for all $i = 0, \dots, 7$.  
    
    An SFC is said to be \textit{closed} or \textit{continuous} if two consecutive points in the image can be orthogonally or diagonally connected \footnote{This property is equivalent to saying that $P$ is continuous in the normed space $(S_l, \left\lVert \cdot \right\rVert_\infty).$}, this is:
    \[
        \left\lVert C^{-1} (i) - C^{-1} (i-1) \right\rVert_\infty = 1, \;\; \text{for } i = 1, \dots 2^{3l}-1.
    \]
\end{definition}

\begin{figure}[htb]
     \centering
     \begin{subfigure}[b]{0.45\textwidth}
         \centering
         \includegraphics[width=0.85\textwidth]{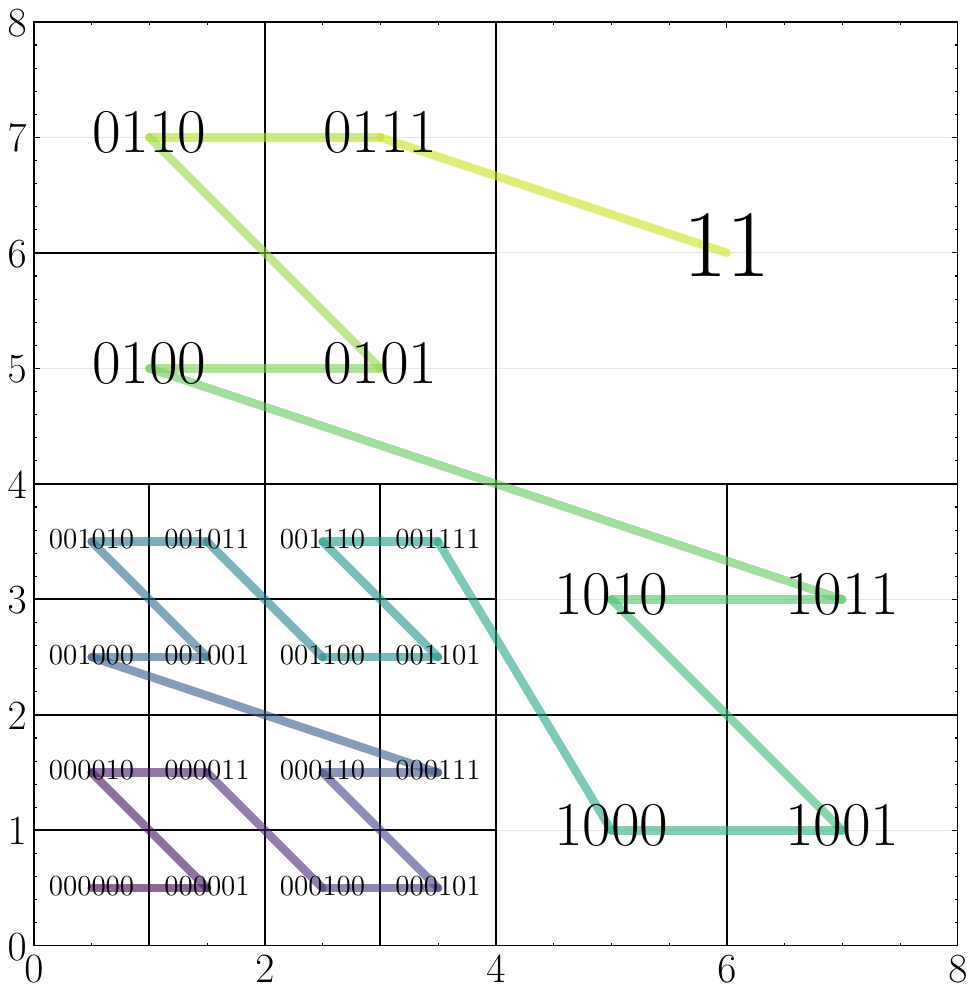}
         \caption{Morton's recursive SFC.}
         \label{subfig:morton-2d}
     \end{subfigure}
     \vspace{1pt}
     \begin{subfigure}[b]{0.45\textwidth}
         \centering
         \includegraphics[width=0.85\textwidth]{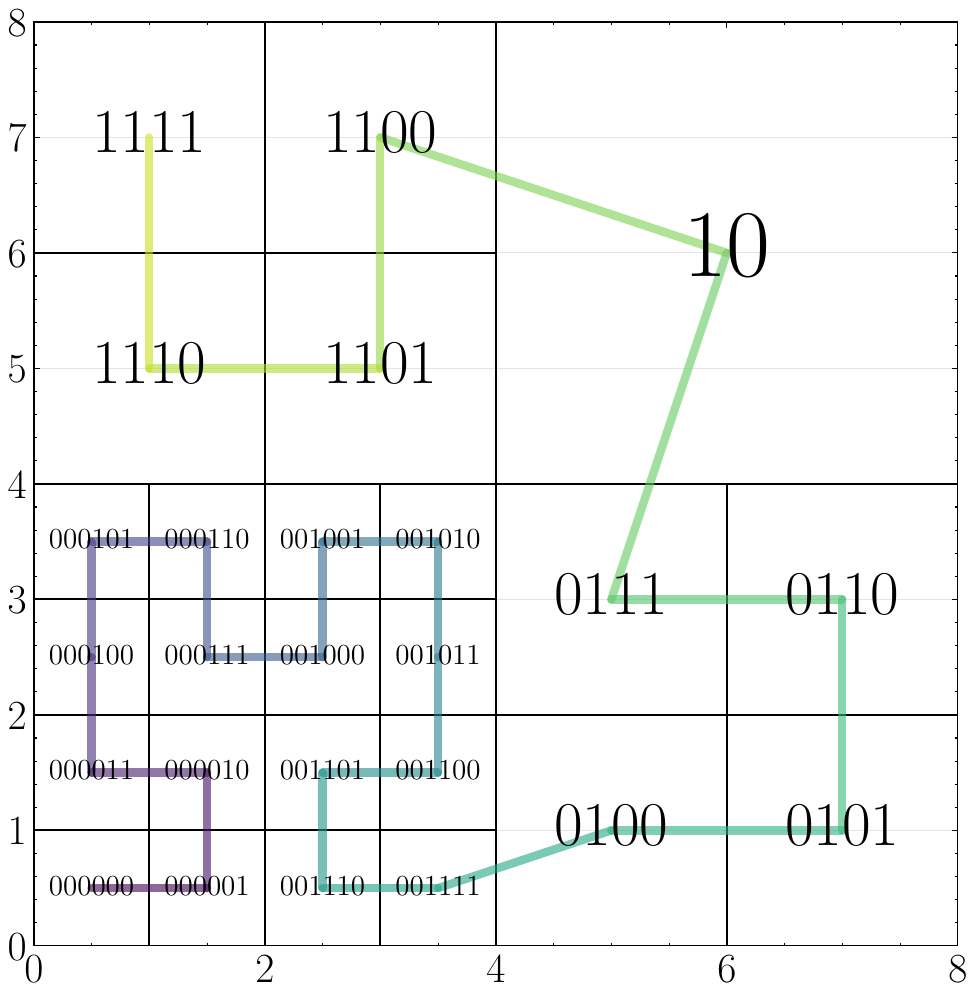}
         \caption{Hilbert's recursive SFC.}
         \label{subfig:hilbert-2d}
     \end{subfigure}
     
     \caption{Two-dimensional SFC visualisations.}
     \label{fig:panel_completo}
\end{figure}

Figures \ref{subfig:morton-2d} and \ref{subfig:hilbert-2d} show the recursive SFCs subdivision space in the two-dimensional case. This process, when carried out in 3 dimensions, forms a dense grid that allows identifying the octants of an \Octree with the codes of each cell.

%%%%%% OLD VERSION: horizontal layour %%%%%%%%%%%%%%
% \begin{figure*}[htpb]
%      \centering
%      \begin{subfigure}[b]{0.45\textwidth}
%          \centering
%          \includegraphics[width=0.85\textwidth]{Figure_2}
%          \caption{Morton's recursive SFC.}
%          \label{subfig:morton-2d}
%      \end{subfigure}
%      \hfill
%      \begin{subfigure}[b]{0.45\textwidth}
%          \centering
%          \includegraphics[width=0.85\textwidth]{Figure_3}
%          \caption{Hilbert's recursive SFC.}
%          \label{subfig:hilbert-2d}
%      \end{subfigure}
     
%      \caption{Two-dimensional SFC visualisations.}
%      \label{fig:panel_completo}
% \end{figure*}
%%%%%%%%%%%%%%%%%%%%%%%%%%%%%%%%%%%%%%%%%%%%%%%%%%%

% ORIGINALMENTE ESTO IBA DESPUES DE HILBERT; PERO ME TIENE MAS SENTIDO AQUI
From these space-filling curves, we can finally reorder the point cloud. The process is as follows:

\begin{enumerate}
    \item For each discretised point $\texttt{p} \in S_L$, consider its encoding using an SFC of level $L$, finding $C(\texttt{p}) = c \in [0, 2^{3L})$. These encodings are stored in the position corresponding to the point in an array $v_c$.
    \item The point cloud $P$ is reordered using the ascending order established by the $v_c$ codes.
\end{enumerate}
As for the computational performance of this process, Step 1 is fully parallelisable, while Step 2 can be efficiently carried out by a parallel \textit{radix sort}. 

\begin{figure*}[htb]
     \centering
     % --- Primera Fila ---
     \begin{subfigure}[b]{0.49\textwidth}
         \centering
         \includegraphics[width=\textwidth]{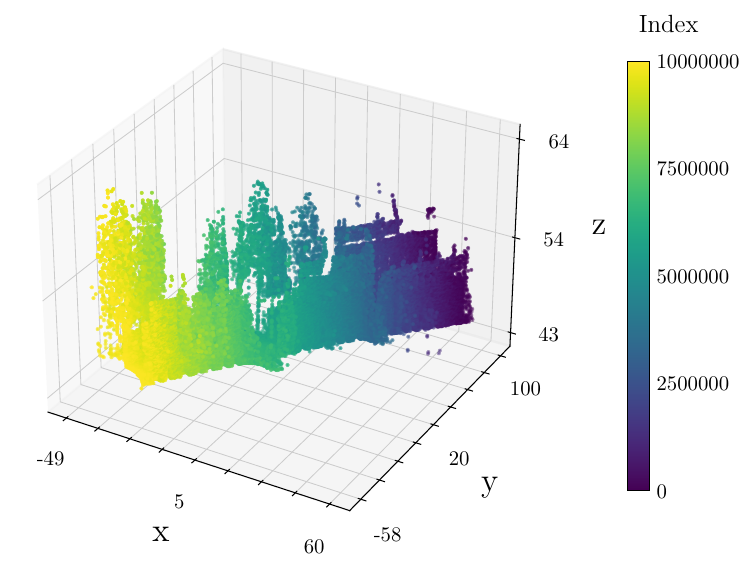}
         \caption{Original order of the points.}
         \label{subfig:lille-0-original}
     \end{subfigure}
     \hfill
     \begin{subfigure}[b]{0.49\textwidth}
         \centering
         \includegraphics[width=\textwidth]{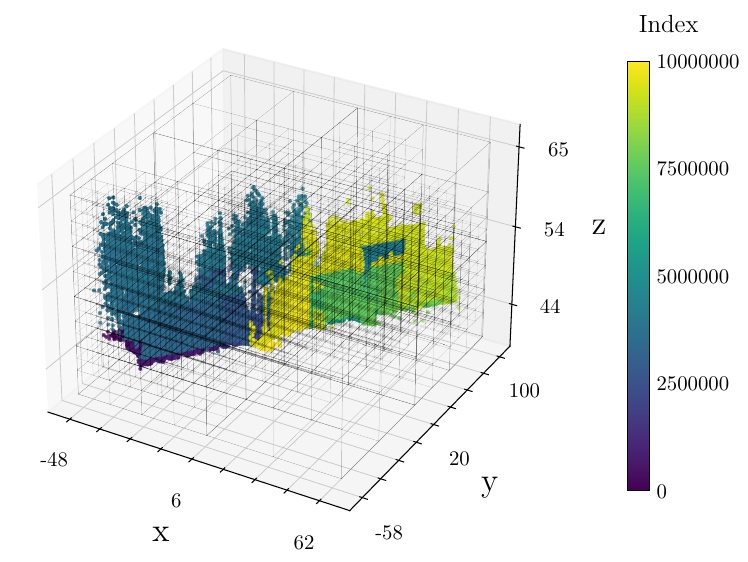}
         \caption{Hilbert reordering with Octree.}
         \label{subfig:lille-0-hilbert-oct}
     \end{subfigure}
     
     \caption{Point cloud reorderings on \textit{Lille\_0} (10M points). The colour scale indicates the order in which the points are stored in memory. Distances are measured in metres.}\label{fig:sfc-reorder-lille-0}
\end{figure*}

In any case, the cloud only needs to be reordered once, and a separate program may perform this task by reading the unordered point cloud and writing it back in the specified order. Figure~\ref{fig:sfc-reorder-lille-0} shows how the reordering affects the storage of the point cloud.

Depending on the topology of the captured 3D data, these jumps between consecutive points in the original order can be quantified. This is reflected in data locality when reordering the cloud.

\subsection{The Morton SFC} % 4.1
The family of recursive SFCs obtained by interlacing the bits of each discretised point is called the Z curve, Lebesgue curve, or Morton curve~\cite{morton1966computer}. For any point $p = (\mathtt{x}, \mathtt{y}, \mathtt{z}) \in S_L$, we have: 
\begin{align*} \label{morton}
    C_M(&\mathtt{x}_0\dots\mathtt{x}_{L-1}, \mathtt{y}_0\dots\mathtt{y}_{L-1}, \mathtt{z}_0\dots\mathtt{z}_{L-1}) =\\ &\mathtt{x}_0\mathtt{y}_0\mathtt{z}_0\dots\mathtt{x}_{L-1}\mathtt{y}_{L-1}\mathtt{z}_{L-1}.
\end{align*}

For instance, let us take $L = 3$, and a point $\mathtt{p} = (\mathcolor{blue}{3}, \mathcolor{orange}{7}, \mathcolor{purple}{2}) = (\mathcolor{blue}{\mathtt{011}}, \mathcolor{orange}{\mathtt{111}}, \mathcolor{purple}{\mathtt{010}}) \in S_3$. Its Morton code will be $C_M(\mathtt{p}) = \mathtt{\mathcolor{blue}{0}\mathcolor{orange}{1}\mathcolor{purple}{0}\mathcolor{blue}{1}\mathcolor{orange}{1}\mathcolor{purple}{1}\mathcolor{blue}{1}\mathcolor{orange}{1}\mathcolor{purple}{0}} = 190 \in [0, 2^9)$. Decoding follows the reverse process: extracting the intertwined bits from the original three integers. Note that there are eight possibilities for the order in which we intertwine each of the bits; that is why we refer to it as a \textit{family} of SFCs. Each of these possibilities corresponds to the order of turns the curve makes. For our purposes, the choice taken is not relevant.

A convenient property of Morton codes is their fast calculation. Using \textit{bitmasks} and look-up tables, it is possible to obtain them with just a few instructions. The \textit{libmorton} library  \cite{libmorton} provides optimised C++ implementations for encoding and decoding. It also leverages vectorisation, using BMI2 and AVX512.

\subsection{The Hilbert SFC} % 4.2
The Hilbert curve is an alternative family of recursive SFCs, obtained through an iterative process. First, we extract and concatenate the bits on a given level $\texttt{x}_l\texttt{y}_l\texttt{z}_l$, as in the Morton curve construction. Then, a series of rotations is performed, dependent on the previous rotation state and the bits extracted in the current step~\cite{lam1994class}. These rotations can be described using a type of Context-Free Grammar (CFG) called \textit{Lindenmayer systems}~\cite{asano1997space,bohm2018novel}.

One advantage of Hilbert codes over Morton codes is that, thanks to the rotations, they are continuous. As depicted in Figure~\ref{subfig:hilbert-2d}, there are no jumps of more than one cell between two consecutive codes when moving from one sub-block to another, unlike Figure~\ref {subfig:morton-2d}. This property is maintained in the three-dimensional case. The family of possible three-dimensional Hilbert curves is large \cite{haverkort2016many}, since those rotations can be performed in many ways. However, only continuity, common to all Hilbert curves, is relevant, as it provides improved spatial locality compared to Morton codes. Hilbert codes are more computationally expensive than Morton codes, as they must be computed iteratively at each subdivision level, and look-up tables cannot be used for direct calculation. In our proposed implementation, the encoding and decoding algorithms described in~\cite{miki2017gothic} are used. 

\section{Enhancing performance through data locality}\label{sec:enhancing-performance-through-data-locality} % 5

Reordering the point cloud using SFCs enables us to introduce the linear \Octree data structure, a variant of the classic pointer-based \Octree. This structure is designed to leverage the improved spatial locality provided by the SFC reordering, allowing for more efficient spatial queries. 

In addition, we present a diagnostic metric based on the distances between the indices of the points in the cloud, which we refer to as \textit{kNN locality histograms}. This metric quantifies the increase in locality achieved by SFC reordering and enables comparison of orderings across datasets; in practice, it can be approximated by computing neighbourhoods on a small subset of points, making it usable as a lightweight predictor for selecting the best SFC ordering prior to full index construction.

\subsection{Linear Octree} \label{sec:linear-oct} % 5.1

In the linear structure, instead of storing the tree explicitly, the structure is synthesised into a set of arrays or maps that contain all the necessary information for the queries. The variant proposed by Keller et al.~\cite{Keller2023} is used in this work due to its efficient and compact representation. The main component of this structure is an array named $leaves$, whose elements are binary numbers addressing a region of space, subdivided by the same process as recursive SFCs, as shown in Figure~\ref{fig:panel_completo}. The process to obtain $leaves$ is as follows:

\begin{enumerate}
    \item Consider the cloud points sorted according to a recursive SFC (Morton or Hilbert). Their codes (also sorted) are needed.
    \item Let $leaves = \{0, 8^L\}$, where $L$ is the maximum level of the \Octree.
    \item Compute the start and end point cloud indices of each of the ranges of points with codes in $[leaves[i], leaves[i+1])$ using binary search on the array of codes.
    \item If the number of points in $[leaves[i],\:leaves[i+1])$ exceeds the given threshold, then this range is subdivided into $8$ equal parts. Return to the previous step until no more subdivisions are necessary.
\end{enumerate}

Another array, called $counts$, stores the number of points in each leaf of the tree. From this array, the range of point indices present in each leaf of the \Octree can be obtained. Next, a second step is required in which the internal part of the tree is computed, internal octant indices are obtained, and depth-first searches can be performed. A detailed description of this tree linking process can be found in the article by Keller et al.~\cite{Keller2023}. Note that this construction can be done almost entirely in parallel and requires only a few heap allocations, resulting in much shorter build times.

Once the internal part of the tree has been obtained, we extend the structure by populating an array \textit{internalRanges}, containing the ranges of cloud indices encompassed by both the leaves and the internal nodes of the linear \Octree.

In this implementation, we can optimise the spatial queries. When an octant is fully contained in the search kernel, all its points can be directly inserted into the result without checking them one by one.

Since it is possible to access points immediately via the indices in \textit{internalRanges}, direct insertion when an octant is entirely contained in the kernel is possible. From this idea, we introduce the optimised algorithm \neighboursPrune as follows:

\begin{enumerate}\label{alg:busquedas-loct}
    \item If the current octant with volume $O$ is a branch, compute the relative position of the node with respect to the kernel $\mathcal{N}$:
    \begin{enumerate}
        \item[1a.] If $O \subseteq \mathcal{N}$, then use \textit{internalRanges} and insert all points sequentially into the result.
        \item[1b.] If $O \cap \mathcal{N} = \emptyset$, then prune this search branch.
        \item[1c.] If $O \cap \mathcal{N} \neq \emptyset$, continue the search.
    \end{enumerate}
    \item If a leaf is reached, check point by point whether or not they belong to the proposed kernel.
    \item Finally, return all the points inside the kernel.
\end{enumerate}

Note that the geometric check to find if the current octant is inside the kernel can be implemented efficiently for our $L_2$ and $L_{\infty}$ norm-based kernels, but it may not generalize to other kernels.

It is also important to consider the format of the result set. Up until now, our search methods have inserted the point objects into an array of point coordinates. However, we can also use an array of index ranges, which greatly reduces its size. This also makes the method scale better as the search radius grows. To get this, we simply replace Step 1a. of \neighboursPrune with the range insertion. This final method will be referred to as \neighboursStruct, as in our implementation, we encapsulate this array of index ranges in a simple structure that provides iterators for posterior access to the neighbourhood points.

Finally, the resulting data structure has the following advantages over the pointer-based \Octree:

\begin{itemize}
    \item Pointer redirection steps that may cause cache misses during deep tree searches are eliminated. The whole linear \Octree structure is contained in a few consecutive memory blocks.
    \item Direct access to the point cloud is preserved in any octant of the tree, whether branch or leaf.
    \item The range of points contained in each octant is known, and thanks to reordering using Morton or Hilbert codes, these points are consecutive in memory.
\end{itemize}

\subsection{Measuring locality via kNN locality histograms} \label{sec:locality-hist} % 5.2

In this section, we analyse the spatial locality we gain during SFC reorderings. For this, we define a dataset-independent locality measure, regardless of the acquisition method or point density, that quantifies how far apart points are in memory in neighbourhood queries.

\begin{definition}
Let $P = \{p_i \in \mathbb{R}^3 : i = 0, \dots, N-1\}$ be a point cloud stored in continuous memory, such that $|i-j| = M\,\lvert\text{addr}(p_i)-\text{addr}(p_j)\rvert$, for a small $M \in \mathbb{N}$. Let $\mathcal{N}_k(i)$ be the set of indices of the $k$ nearest neighbours of $p_i$, and $D_k(i, d) = \left|\{ j \in \mathcal{N}_k(i) : |i-j| = d\}\right| \in \{0, 1, 2\}$ be the amount of points in $\mathcal{N}_k(i)$ at a distance of $d$ to $i$. Then, we define the \textit{kNN locality histogram} $H_k: \{0, \dots, N-1\} \rightarrow \mathbb{N}$ as
$$
H_k(d) = \sum_{i=0}^{N-1} D_k(i, d).
$$
\end{definition}

There are two immediate properties that help to understand what this definition means:
\begin{itemize}
\item $H_k(0) = N$, since every kNN search will find its own centre at distance 0, and thus $i \in \mathcal{N}_k(i)$ and $D_k(i, d) = 1$.
\item $\sum_{d=0}^{N-1} H_k(d) = kN$, as every search finds exactly $k$ points (assuming $k \leq N$), so we get $kN$ points in total.
\end{itemize}

Also note that we define $H_k$ based on kNN searches because fixed-radius queries depend on local cloud density. This dependence makes it harder to find an appropriate radius for analysis and to compare results across multiple clouds.

The problem of improving locality can now be reformulated as making this histogram as left-skewed as possible. To measure skewness, we will use the Fisher-Pearson coefficient:
$$
G_1 = E\left[\left(\frac{d-\mu}{\sigma}\right)^3\right].
$$

With $\mu$ and $\sigma$ being the mean and standard deviation of the histogram, respectively. Although it would be ideal to have a tight lower bound $H_k(d) \in \Omega(k)$ with a small constant, this is unattainable due to the impossibility of finding a continuous reverse SFC map $[0, 2^{3l}) \to S_l$. All SFCs eventually exhibit large gaps when switching between sub-blocks of code.

Since computing $H_k$ exactly requires finding all neighbourhoods, it is intended as a diagnostic metric rather than a runtime tool. Its practical value lies in the assumption that by computing $H_k$ on a small random subset of query points (e.g., 1\% of the cloud), the skewness of each candidate SFC ordering can be estimated at low cost and select the one with the highest skewness prior to full index construction. The spatial symmetry of SFCs and the density-independence of kNN distances make this subset approximation reliable across datasets with varying acquisition patterns.

\section{Experimental configuration}\label{sec:exp_env} % 6

The experiments were conducted on a system equipped with four Intel(R) Xeon(R) E5-4620 v4 processors (Broadwell architecture) running at 2.10 GHz. Each processor features 10 physical cores, for a total of 40, and supports 251 GiB of RAM. The software environment was managed under Alma Linux 8.6 (Sky Tiger) with Kernel 4.18.0. Our linear \Octree implementation is available as a C++ library \footnote{\scriptsize \faGithub \space \url{https://github.com/linear-octree/linear-octree}} and our benchmarking code can be found at \footnote{\scriptsize \faGithub \space \url{https://github.com/linear-octree/linoctree-benchmark}}. Our code was compiled using GCC 12.1.1, with the -O2 optimisation level enabled.

In Table~\ref{tab:algorithms}, the benchmarked implementations are summarised. The \ptrOctree{} structure is a simple pointer-based \Octree implementation, serving as a baseline for the optimised linear \Octree described in Section~\ref {sec:linear-oct}.

\begin{table}[htb]
\setlength{\tabcolsep}{2.5pt}
\caption{Spatial data structures evaluated in this study and the corresponding fixed-radius and kNN search methods.}\label{tab:algorithms}
\centering
%\footnotesize
\scriptsize
\begin{tabular}{llll}
\toprule
Structure & Type & Fixed-radius method & kNN method \\
\midrule
\ptrOctree & Pointer-based \Octree & \neighboursPtr & --- \\
\addlinespace

\linOctree\xspace \textbf{(ours)} & Linear \Octree & \neighboursLin & \knnLinOct \\
& & \neighboursPrune & \\
& & \neighboursStruct & \\
\addlinespace

\unibnOctreeLib ~\cite{behley2015efficient} & Pointer-based \Octree & \neighboursUnibn & --- \\
\addlinespace

\pclOctree ~\cite{PCL2011} & Pointer-based \Octree & \neighboursPCLOct  & \knnPCLOct \\

\pclKDTree ~\cite{PCL2011} & \KDTree & \neighboursPCLKD   & \knnPCLKD \\

\nanoflannKDTree ~\cite{Nanoflann2014}   & \KDTree & \neighboursNano & \knnNano \\

\picotreeKDTree ~\cite{picotree} & \KDTree & \neighboursPico & \knnPico \\

\bottomrule
\end{tabular}
\end{table}

To evaluate the impact of the different reorderings on point neighbourhood searching using \Octrees, our experiments are performed on diverse well-known datasets of LiDAR point clouds:

\begin{itemize}
    \item \textit{Paris-Lille-3D} \cite{roynard2018paris}: This dataset maps two streets in Paris and Lille. It has a high density and was captured using terrestrial LiDAR technology, with a sensor mounted on a vehicle. Segments of 10 million points such as \textit{Lille\_0} or \textit{ParLux\_6} as well as the full clouds \textit{Lille} and \textit{ParLux} will be used.
    \item \textit{DALES} \cite{varney2020dales}: This is an aerial LiDAR dataset divided into disjoint tiles of approximately 10 million points. Its density is lower, so, on average, fewer points will be found in the neighbourhoods. We perform our experiments in tiles such as \textit{5080\_54400}.
    \item \textit{Semantic3D} \cite{hackel2017isprs}: This is a high-density LiDAR dataset captured with a fixed sensor on a tripod. Specifically, clouds \textit{bildstein\_station1} (30M points) and \textit{sg27} (430M points) are used in our analysis.
    \item \textit{Speulderbos} \cite{brede2020speulderbos}: This dataset was obtained using a LiDAR sensor mounted on a drone and on a ground tripod. It has the highest density, consisting of approximately 721 million points.
\end{itemize}

All \Octree structures are configured with a maximum leaf capacity of $N_{max} = 128$ points.
This choice is explained in more detail in Section~\ref{sec:memory-footprint-and-build-times}.

\section{Results and discussion}\label{sec:results} % 7

In this section, we compare the performance of our linear Octree with the classic pointer-based Octree and with several widely used state-of-the-art libraries for point cloud processing workflows. Also, we carefully measure the impact of point reordering on memory access times. Finally, we evaluate the efficiency of the parallel implementation of the query algorithms.

\subsection{Comparison between linear and pointer-based Octrees}\label{sec:locality-benchmarks} % 7.1

\begin{figure*}[hbt]
    \centering  
    \includegraphics[width=0.8\linewidth]{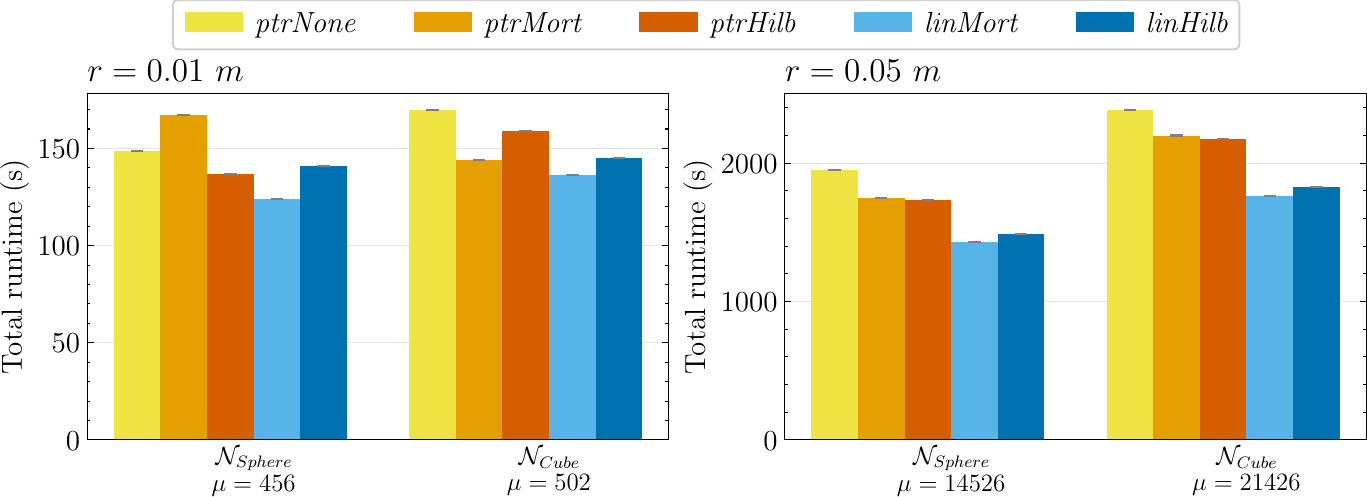}
    \caption{Full search runtimes for different kernels and radii on cloud \textit{sg27}.}
    \label{fig:sg27-full}
\end{figure*}

\begin{figure*}[hbt]
    \centering  
    \includegraphics[width=0.8\linewidth]{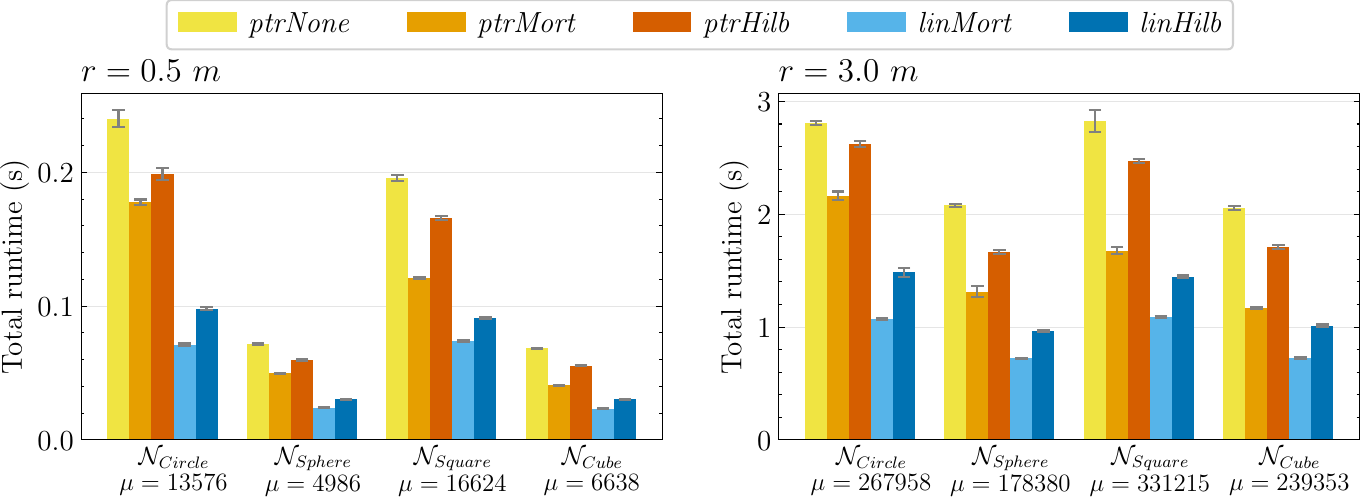}
    \caption{Random search runtimes for different kernels and radii on cloud \textit{ParLux\_6}.}
    \label{fig:paris-lux-random}
\end{figure*}

In this section, we analyse the performance of our \Octrees for fixed-radius searches. We consider runtimes for parallel cloud searches over a set of query centres $v_s$, distinguishing between \textit{full neighbourhood searches} (over $v_s = P$) and \textit{random neighbourhood searches} (over a random subset $v_s \subset P$, $|v_s| = 5000$). Full searches measure finding all-to-all neighbourhoods, while random searches measure single-access performance. Note that we employ larger radii for random queries to better evaluate single-access performance under high workload, and smaller radii for full queries. The latter already incurs substantially higher cost, and larger radii would lead to excessive execution times without affecting the qualitative conclusions.

Full point cloud searches behave essentially differently from random ones, as each thread gets a consecutive chunk of centres. This makes SFC reordering more effective, since the neighbours of previously computed points are often in each core's local cache.

To compare both SFC reordering and \Octree implementation, we use the five configurations in Table \ref{tab:conf-octrees-sfc}. For both methods \neighboursPtr and \neighboursLin, we apply different SFC reorderings. Note that \linNone is missing since the linear \Octree relies on SFC reordering to be built.

\begin{table}[htb]
    \centering
    \caption{\Octree and SFC configurations used in Section \ref{sec:locality-benchmarks}.}\label{tab:conf-octrees-sfc}
    %\footnotesize
    \scriptsize
    \begin{tabular}{lll}
        \toprule
        Method & SFC & Configuration \\
        \midrule

        \multirow[t]{3}{*}{\neighboursPtr}
        & None    & \ptrNone \\
        & Morton  & \ptrMort \\
        & Hilbert & \ptrHilb \\

        \addlinespace

        \multirow[t]{2}{*}{\neighboursLin}
        & Morton  & \linMort \\
        & Hilbert & \linHilb \\

        \bottomrule
    \end{tabular}
\end{table}

Figures \ref{fig:paris-lux-random} and \ref{fig:sg27-full} show the accumulated runtimes for random searches on cloud \textit{ParLux\_6} and full searches on cloud \textit{sg27}. For each radius, bar charts show the measures for each kernel. The average number of points found, $\mu$, is annotated below each run. We observed qualitatively similar trends on the other clouds for these experimental conditions.

Note that the linear \Octree performs better when $\mu$ is large, even without the pruning optimisations. SFC reorderings also improve runtimes in most cases for the pointer-based \Octree. On sg27, it is harder to see how Morton and Hilbert reorderings improve the results, since as will be shown in next section, original locality was already high, and cache misses are not reduced significantly.

\begin{figure}[htb]
    \centering  
    \includegraphics[width=0.7\linewidth]{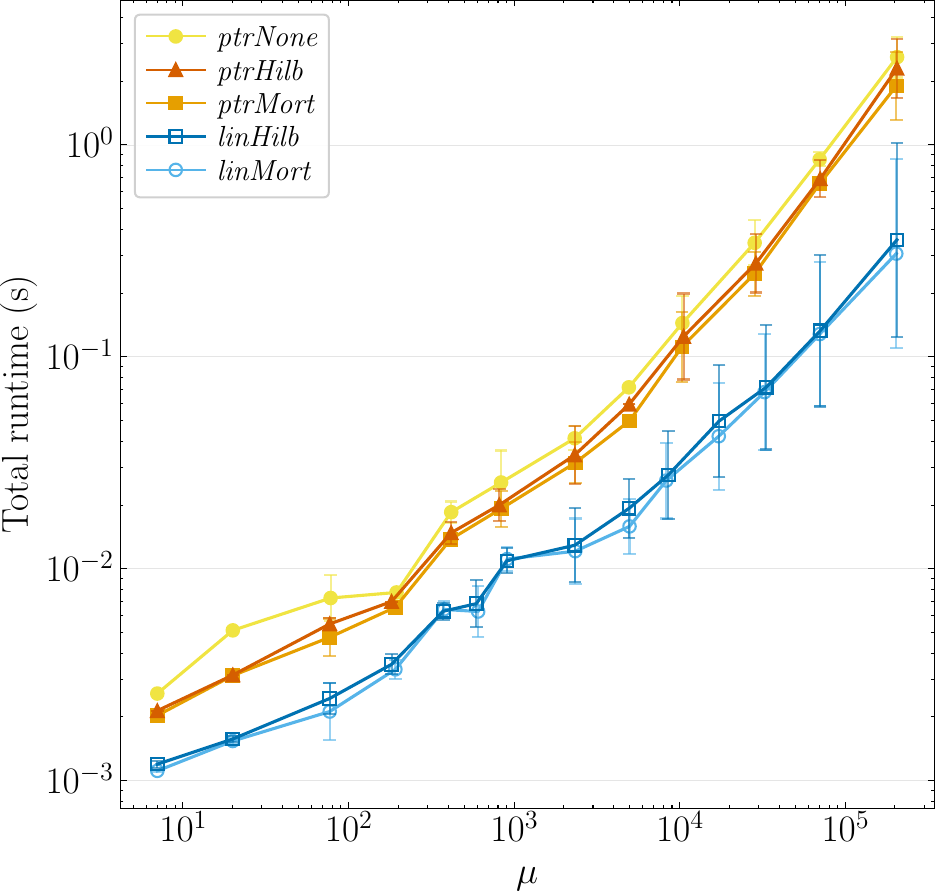}
    \caption{Search runtimes against $\mu$ across different kernels, radii and clouds for basic implementation on pointer-based and linear \Octrees and different SFC reorderings.}
    \label{fig:loglog-random}
\end{figure}

To get a more general picture, note that $\mu$ is the main factor regarding the scalability of the methods across multiple clouds, as the same radius $r$ can return different amounts of points depending on cloud density. For every dataset and every tested combination of radius and kernel, we present in Figure~\ref{fig:loglog-random} how each combination of SFC and \Octree performs on random searches. The plot is in logarithmic scale on both axes, representing $\mu$ and total runtime, which visually compresses differences between methods. SFCs marginally improve runtime on the pointer-based \Octree without changing the base algorithm. The linear \Octree optimised memory layout further reduces the runtime, while essentially using the same algorithm as the pointer-based version.

\begin{figure}[htb]
    \centering  
    \includegraphics[width=0.7\linewidth]{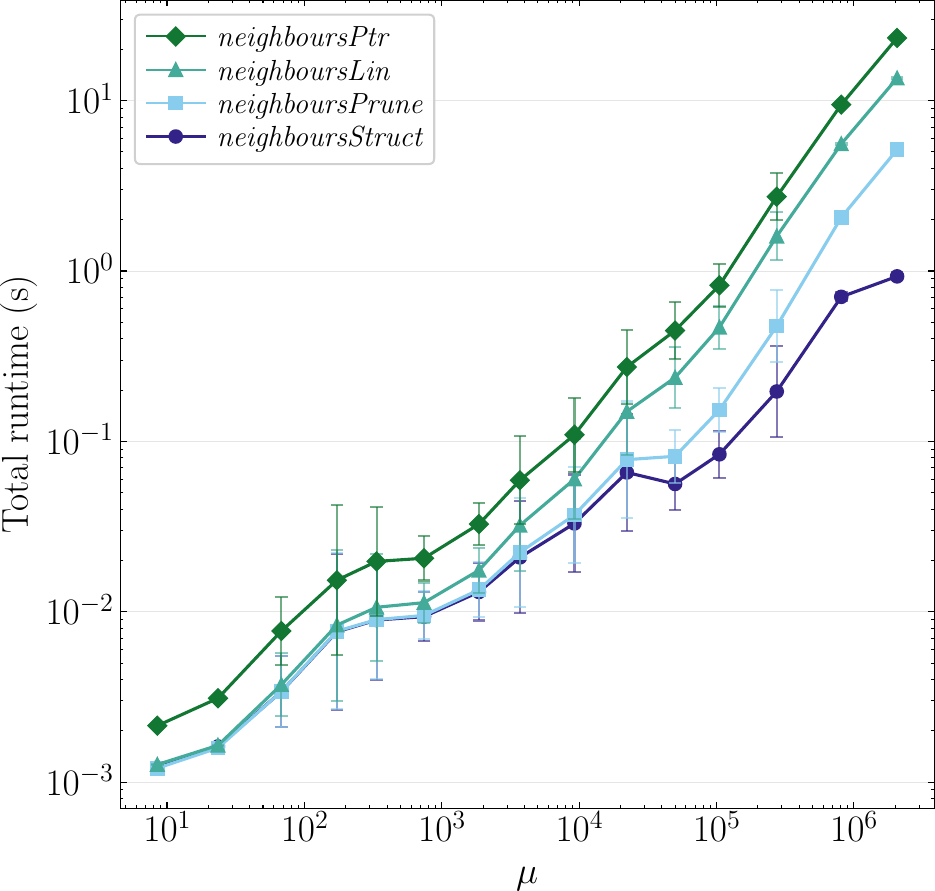}
    \caption{Search runtimes against $\mu$ across different kernels, radii and clouds for each algorithm developed and Hilbert reordering.}
    \label{fig:loglog-random-algos}
\end{figure}

The difference between Morton and Hilbert SFCs is hard to discern, as both curves yield similar results in most configurations. Theoretically, the Hilbert curve exhibits better locality, as we will later analyse in Section \ref{sec:sfc-benchmarks}. However, the small reduction in cache misses observed in kNN searches can be overshadowed by the variance in point distribution and density during fixed-radius searches.

The effect of the optimised methods \neighboursPrune and \neighboursStruct is analysed in Table \ref{tab:results-algos}. In this table, we compare the performance of the search methods for a single radius and the $\kernel{Sphere}$ kernel. Figure \ref{fig:loglog-random-algos} shows the general situation for random searches across all tested datasets, kernels, and radii. As expected, the scalability of methods \neighboursPrune and \neighboursStruct greatly surpasses the basic algorithm, achieving an improvement of over one order of magnitude for big neighbourhoods. To understand the hardware-level mechanisms driving these improvements, we next examine how SFC reordering enhances spatial locality and reduces cache misses.

\begin{table}[htb]
    \centering
    \caption{Full search runtimes for our search methods with the $\kernel{Sphere}$ kernel, after Hilbert reordering.}\label{tab:results-algos}
    \footnotesize
    \begin{tabular}{llll}
    \toprule
    Cloud & $r$ (m) & Search method & Runtime (s) \\
    \midrule

    \textit{Lille\_0} & 3.0 & \dotneighboursPtr\neighboursPtr & 265.63 \\
     &   & \dotneighbours\neighboursLin & 159.04 \\
     &   & \dotneighboursPrune\neighboursPrune & 100.39 \\
     &   & \dotneighboursStruct\neighboursStruct & \textbf{33.92} \\

    \addlinespace

    \textit{ParLux\_6} & 3.0 & \dotneighboursPtr\neighboursPtr & 1051.45 \\
     &   & \dotneighbours\neighboursLin & 696.88 \\
     &   & \dotneighboursPrune\neighboursPrune & 352.61 \\
     &   & \dotneighboursStruct\neighboursStruct & \textbf{53.57} \\

    \addlinespace

    \textit{5080\_54400} & 10.0 & \dotneighboursPtr\neighboursPtr & 46.84 \\
     &   & \dotneighbours\neighboursLin & 42.37 \\
     &   & \dotneighboursPrune\neighboursPrune & 36.48 \\
     &   & \dotneighboursStruct\neighboursStruct & \textbf{25.78} \\

    \addlinespace

    \textit{sg27} & 0.05 & \dotneighboursPtr\neighboursPtr & 1639.95 \\
     &   & \dotneighbours\neighboursLin & 1321.82 \\
     &   & \dotneighboursPrune\neighboursPrune & 976.02 \\
     &   & \dotneighboursStruct\neighboursStruct & \textbf{413.16} \\

    \addlinespace

    \textit{Speulderbos} & 0.25 & \dotneighboursPtr\neighboursPtr & 1126.93 \\
     &   & \dotneighbours\neighboursLin & 936.02 \\
     &   & \dotneighboursPrune\neighboursPrune & 805.68 \\
     &   & \dotneighboursStruct\neighboursStruct & \textbf{476.36} \\

    \bottomrule
    \end{tabular}
\end{table}

\subsection{Impact of SFC reordering on kNN locality histograms and cache misses} \label{sec:sfc-benchmarks} % 7.2

\begin{figure*}[htb]
    \centering
    \begin{subfigure}{0.48\textwidth}
        \centering
        \includegraphics[width=0.85\linewidth]{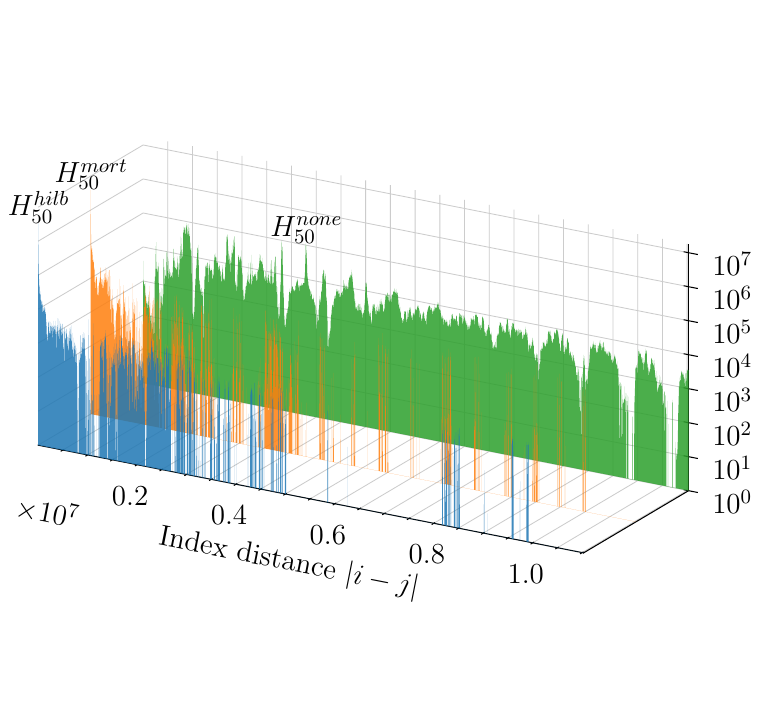}
        %\vspace{-40pt} % hack because figure has a very big margin
        \vspace{-25pt} % hack because figure has a very big margin
        \caption{\textit{5080\_54400}}
         \label{fig:locality-hist-5080-54400}
    \end{subfigure}
    \hfill
    \begin{subfigure}{0.48\textwidth}
        \centering
        \includegraphics[width=0.85\linewidth]{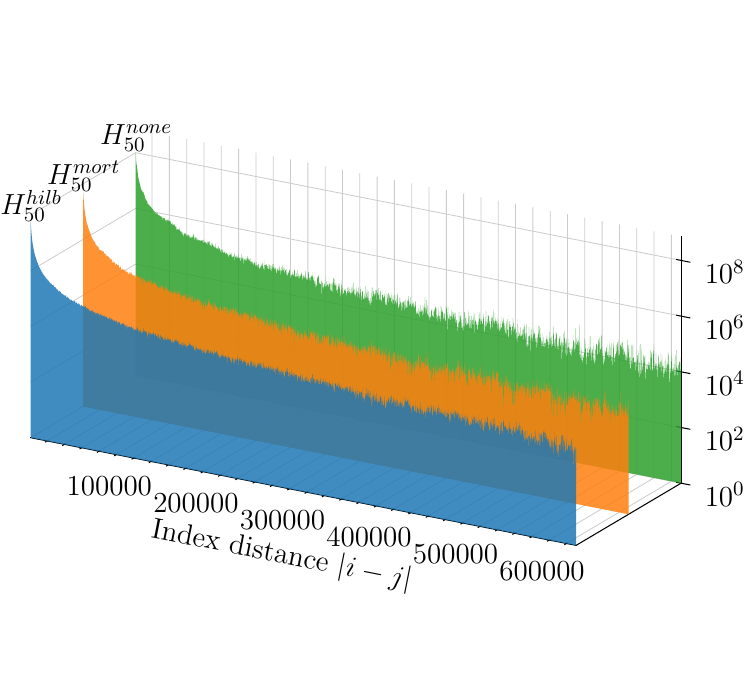}
        %\vspace{-40pt} % hack because figure has a very big margin
        \vspace{-25pt} % hack because figure has a very big margin
        \caption{\textit{sg27}}
        \label{fig:locality-hist-sg27}
    \end{subfigure}
    \caption{kNN locality histograms.}
    \label{fig:locality-hist}
\end{figure*}

We now measure how SFCs affect the kNN locality histogram defined in Section~\ref{sec:locality-hist}. As an illustrative example, in Figure~\ref{fig:locality-hist} we display the results for Morton and Hilbert SFCs on clouds 
\textit{5080\_54400} and \textit{sg27} using $k = 50$, as representative examples of different densities and acquisition methods.

\begin{table*}[htb]
\centering
\caption{Average L1d cache misses per search centre on kNN searches ($k = 50$) and associated spatial locality metrics from $H_{50}$. Lower ($\downarrow$) is better for L1d misses and the quantiles, higher ($\uparrow$) is better for $G_1$. $G_1$ (1\%) and $G_1$ (0.1\%) are skewness estimates computed on random subsets of the respective size.}\label{tab:locality}
%\footnotesize
\scriptsize
\begin{tabular}{lllllllll}
\toprule
Cloud & SFC & L1d Misses per point $\downarrow$
& \textbf{$Q_1$} $\downarrow$
& \textbf{$Q_2$} $\downarrow$
& \textbf{$Q_3$} $\downarrow$
& $G_1 \uparrow$
& $G_1$ (1\%) $\uparrow$
& $G_1$ (0.1\%) $\uparrow$ \\
\midrule

\textit{Lille\_0}
& None    & 15.77 & $1.67 \cdot 10^6$ & $3.33 \cdot 10^6$ & $6.02 \cdot 10^6$ & 0.56 & 0.56 & 0.56 \\
& Morton  & 5.04  & 12 & 57 & 348 & 16.32 & 16.42 & \textbf{17.06} \\
& Hilbert & \textbf{4.68} & \textbf{10} & \textbf{42} & \textbf{266} & \textbf{19.29} & \textbf{18.39} & 16.73 \\

\addlinespace

\textit{5080\_54400}
& None    & 19.29 & $1.67 \cdot 10^6$ & $3.33 \cdot 10^6$ & $6.02 \cdot 10^6$ & 0.56 & 0.56 & 0.56 \\
& Morton  & 6.04  & 12 & 57 & 348 & 16.32 & 16.42 & \textbf{17.06} \\
& Hilbert & \textbf{5.53} & \textbf{10} & \textbf{42} & \textbf{266} & \textbf{19.29} & \textbf{18.39} & 16.73 \\

\addlinespace

\textit{sg27}
& None    & 8.56 & \textbf{8} & \textbf{27} & \textbf{117} & 55.96 & 55.88 & 54.10 \\
& Morton  & 7.06 & 10 & 55 & 270 & 58.45 & 58.00 & 56.79 \\
& Hilbert & \textbf{6.89} & 9 & 40 & 205 & \textbf{84.66} & \textbf{84.07} & \textbf{83.63} \\

\bottomrule
\end{tabular}
\end{table*}

Note that both SFCs provide similar results, making $H$ more skewed towards lower values and smoothing out the distances. In \textit{5080\_54400}, the original order generates spikes at high distances due to the way the LiDAR points were obtained,  via at least 4 passes through each region \cite{varney2020dales}. Most of those spikes get removed after the reorderings, massively improving locality. In contrast, other clouds like \textit{sg27} already provide a good initial ordering, making SFC reordering less impactful.

From these locality gains, we expect a reduction in cache misses during parallel searches. We have measured L1d, L2d and L3 misses using PAPI, a portable interface to hardware performance counters \cite{papi}. Table~\ref{tab:locality} shows the average number of L1d misses per search centre for kNN searches using \nanoflann, along with the quantiles and the skewness metric $G_1$ for the associated kNN locality histograms. As hinted before, calculating $G_1$ with a small sample of points yields a good approximation, making this metric useful for reasoning which of the two encodings is more cache-friendly before running full cloud searches. Miss counts for L2d and L3 misses are similar. Both SFCs eliminate a significant amount of misses, from $20\%$ to $75\%$, with Hilbert's being slightly better on all tested point clouds. As expected, clouds with smaller locality improvements, such as \textit{sg27} also exhibit smaller reductions in cache misses compared to the unordered case. While these locality gains explain the internal speedups of our Octree variants, we also evaluate our proposal against other state-of-the-art spatial search libraries in the following section.

\subsection{Comparison with other search methods} \label{sec:comparison-libraries} % 7.3

\begin{figure}[htb]
    \centering  
    \includegraphics[width=0.8\linewidth]{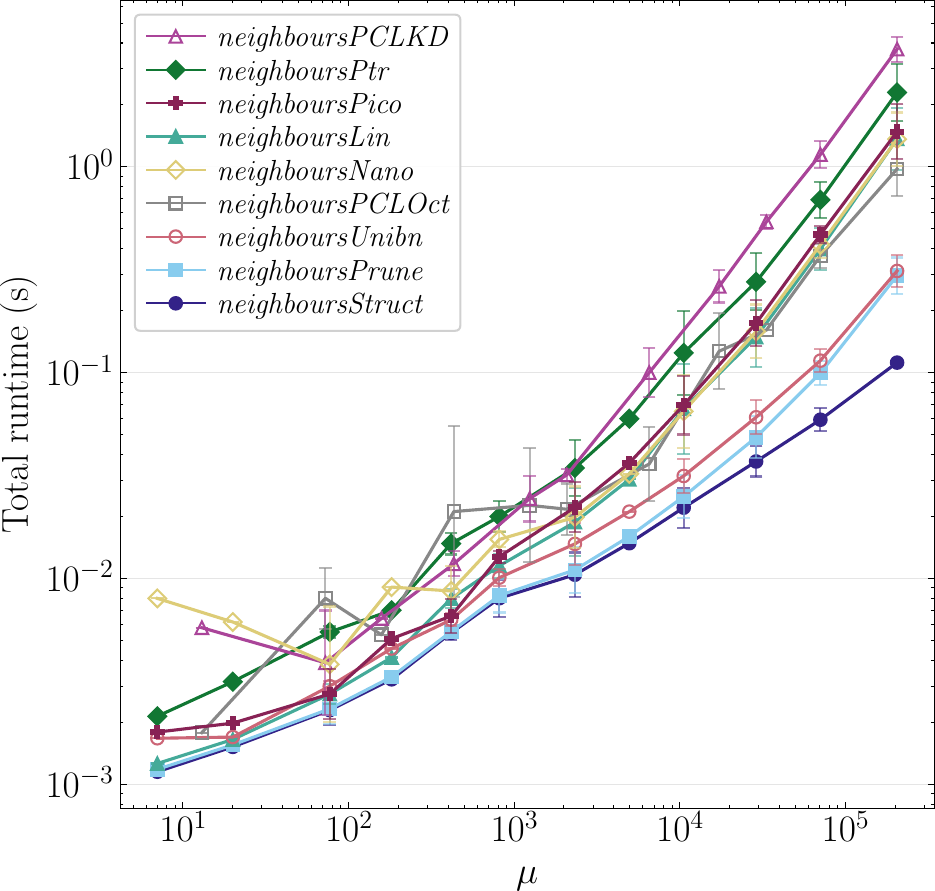}
    \caption{Search runtimes for random searches against $\mu$ across different radii and clouds for each structure tested and kernel $\kernel{Sphere}$, after Hilbert reordering.}
    \label{fig:loglog-random-structures}
\end{figure}

We also compare our methods with state-of-the-art libraries such as \unibnOctreeLib~\cite{behley2015efficient}, PCL's \Octree and \KDTree~\cite{PCL2011}, \nanoflann~\cite{Nanoflann2014}, and \picotree~\cite{picotree}. As not all libraries efficiently support all search kernels, we focus on $\kernel{Sphere}$ on this test, as it is the most commonly supported query shape across the evaluated libraries. Figure~\ref{fig:loglog-random-structures} shows the results for random searches. We report average runtimes over 5 independent executions.

Note that our most efficient implementation, \neighboursStruct, performs well all over the board, especially for queries involving a large number of points (with $\mu >10^3$). The optimised \neighboursPrune implementation, which returns points as an array of coordinates, performs similarly to \unibnOctreeLib. \picotree, \nanoflann\xspace and PCL's \Octree follow next, and lastly we have our basic implementations \neighboursLin and \neighboursPtr along with PCL's \KDTree. 

The performance advantage of \neighboursStruct over coordinate-returning libraries has two distinct sources: (i) the algorithmic gains of the linear \Octree (early termination via \neighboursPrune, contiguous memory layout, and SFC-induced cache locality), and (ii) the reduced output allocation cost of returning index ranges instead of 3D coordinates. To isolate the algorithmic contribution, \neighboursPrune, which shares the same search logic but returns an array of coordinates, can be taken as the fairer comparison point against external libraries; its results are also shown in Figure~\ref{fig:loglog-random-structures} and confirm that the linear \Octree is competitive independently of the output format. Note that the range-based output of \neighboursStruct is directly usable in downstream pipelines that iterate over point indices, which is the common case in segmentation, feature extraction, and normal estimation workflows.

From these results, we conclude that for big enough radii, our fixed-radius search method performs better on 3D LiDAR data than the other \Octree and \KDTree implementations.

\begin{figure*}[htb]
    \centering
    \begin{subfigure}{0.48\linewidth}
        \centering
        \includegraphics[width=0.75\linewidth]{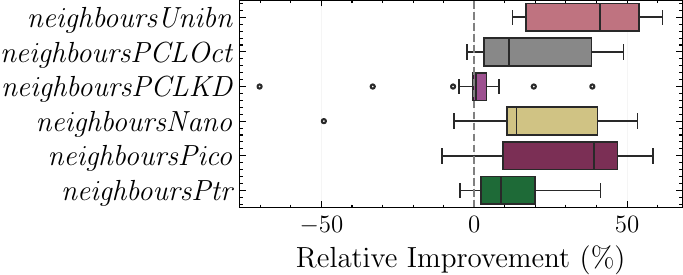}
        \caption{Random searches.}
        \label{subfig:none-vs-hilb-random}
    \end{subfigure}
    \hfill
    \begin{subfigure}{0.48\linewidth}
        \centering
        \includegraphics[width=0.75\linewidth]{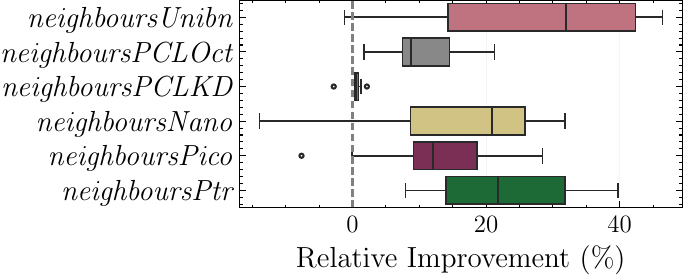}
        \caption{Full searches.}
        \label{subfig:none-vs-hilb-full}
    \end{subfigure}
    \caption{Relative runtime of multiple fixed-radius neighbour search algorithms. The relative runtime of the Hilbert reordered cloud is compared against a no-reordered baseline. The linear \Octree is not included since it can not be built on unordered clouds.}
    \label{fig:none-vs-hilb}
\end{figure*}

\begin{figure*}[htpb]
    \centering
    \begin{subfigure}{0.48\linewidth}
        \centering
        \includegraphics[width=0.75\linewidth]{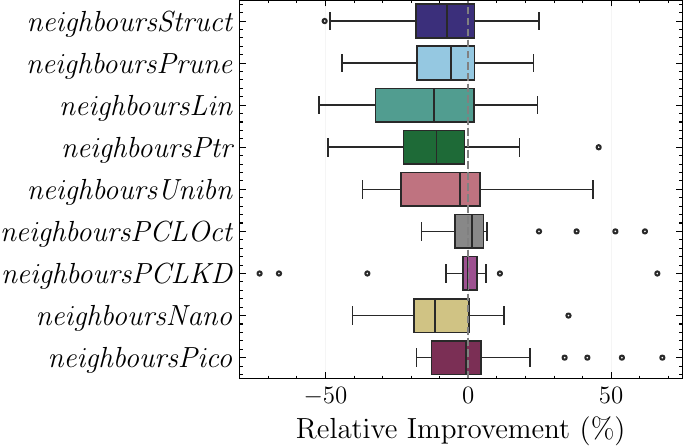}
        \caption{Random searches.}
        \label{subfig:mort-vs-hilb-random}
    \end{subfigure}
    \hfill
    \begin{subfigure}{0.48\linewidth}
        \centering
        \includegraphics[width=0.75\linewidth]{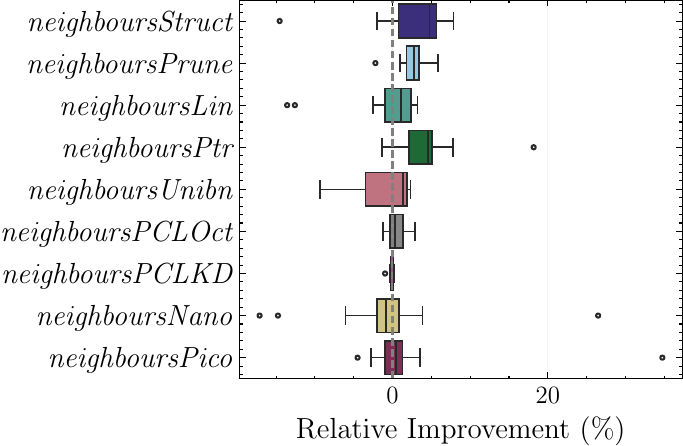}
        \caption{Full searches.}
        \label{subfig:mort-vs-hilb-full}
    \end{subfigure}
    \caption{Relative runtime of multiple fixed-radius neighbour search algorithms. The relative runtime of the Hilbert reordered cloud is compared with a Morton reordered baseline.}\label{fig:mort-vs-hilb}
\end{figure*}

Finally, we compare how each structure performs under different SFC encodings (original order, Morton, and Hilbert). Figure~\ref{fig:none-vs-hilb} provides a comparison between no reordering and Hilbert reordering, and Figure~\ref{fig:mort-vs-hilb} shows another for Morton against Hilbert reordering. 

Note that SFC reorderings enhance all sorts of spatial query algorithms. The Hilbert SFC typically reduces runtime by $10\%$ to $50\%$ for most tested algorithms and configurations, across both random and full searches, compared to an unordered cloud. On these plots, we do not show linear \Octree results, as that structure can not be constructed without a prior SFC reordering.

The difference between Morton and Hilbert SFC is more subtle. For random searches in Figure~\ref{subfig:mort-vs-hilb-random}, Morton SFC seems more beneficial in more clouds than Hilbert SFC. However, we may attribute this effect to two factors: first, the searches operate on relatively small subsets of points, thereby limiting the working set size and mitigating the impact of cache misses; second, the random selection of query centres limits the usefulness of Hilbert SFC continuity. As a result, the theoretical spatial locality offered by the Hilbert curve is less significant in this context. For full searches in Figure~\ref{subfig:mort-vs-hilb-full}, we observed a slight improvement, especially when running the algorithms from our linear \Octree.

\begin{figure}[htb]
    \centering
    \begin{subfigure}{0.8\linewidth}
        \centering
        \includegraphics[width=\linewidth]{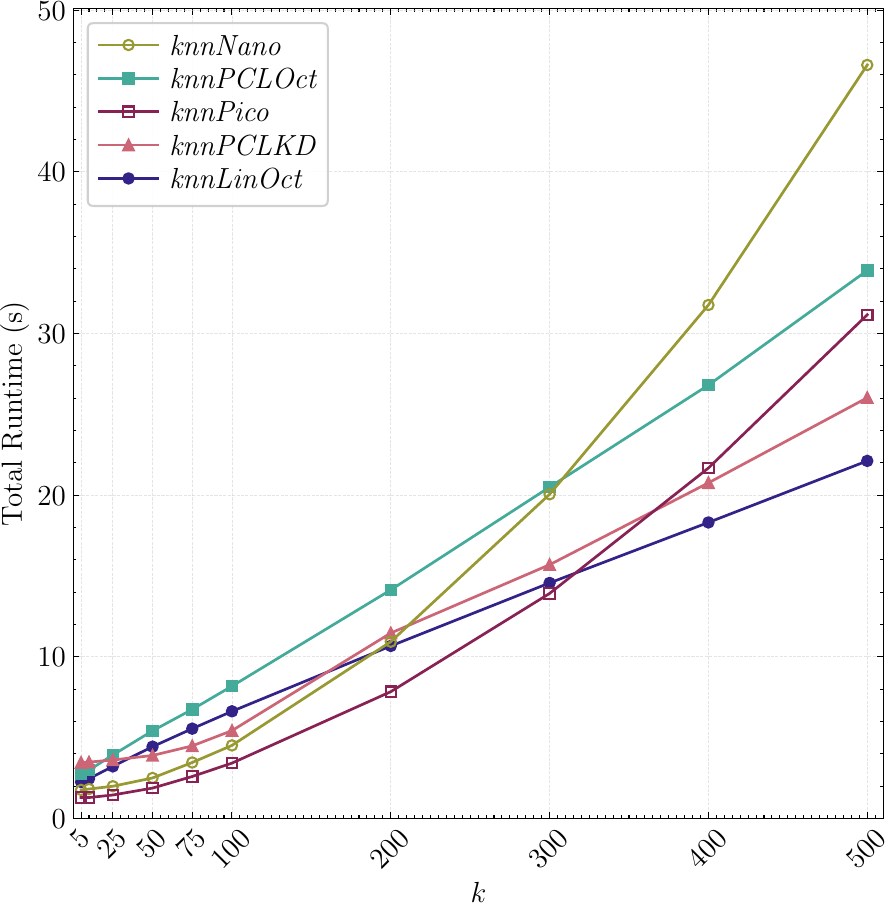}
        \caption{\textit{ParLux\_6}, full.}
        \label{subfig:knn-parislux6-full}
    \end{subfigure}    
    \vspace{1pt}
    \begin{subfigure}{0.8\linewidth}
        \centering
        \includegraphics[width=\linewidth]{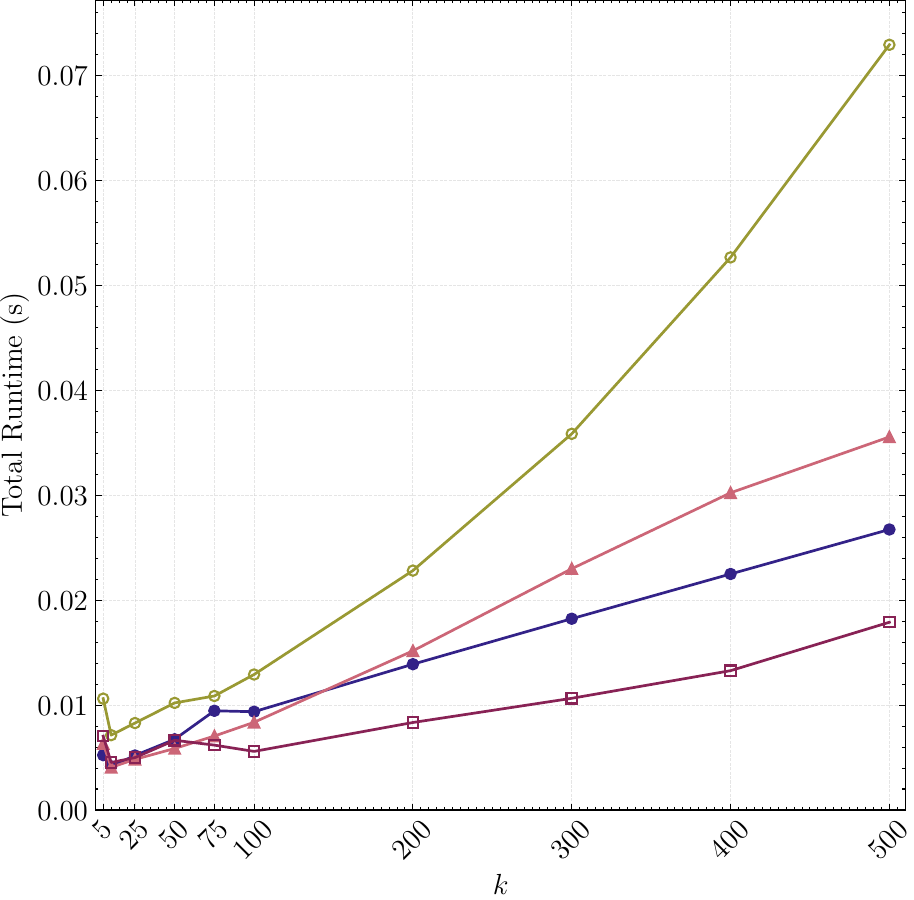}
        \caption{\textit{sg27}, random.}
        \label{subfig:knn-sg27-random}
    \end{subfigure}
    \caption{kNN runtime comparisons, after Hilbert reordering.}\label{fig:knn-bench}
\end{figure}

In the case of kNN searches, the \textit{linOctreeKNN} implementation was considered varying the value of $k$. Note that $k$ is independent of cloud density, making it more suitable for direct comparison between different datasets. We benchmark our linear \Octree against the previously mentioned structures after Hilbert reordering of the evaluated clouds. Figure~\ref{fig:knn-bench} shows some results for random and full cloud kNN searches. We omit PCL's \Octree in Figure~\ref{subfig:knn-sg27-random} because it is much slower than the others, which would otherwise complicate the visualisation of the data. ParLux\_6 and sg27 are used as examples, as similar traces were observed in the rest of the evaluated datasets.

For random searches, \picotree outperforms the other implementations over all tested $k$ values, followed by our own implementation \textit{linOctreeKNN} and PCL and nanoflann \KDTrees. For full searches, our implementation performs better in some of the datasets when $k$ is bigger than a certain threshold (typically on the $300-500$ range). This performance gap for random kNN at small $k$ (typically $k < 300$) is expected: \KDTree-based methods such as \picotree use axis-aligned splits that tightly bound the search volume, minimising the number of nodes visited for small neighbourhoods. In contrast, the \Octree decomposition produces cubic cells that are less adaptive to the local point distribution, and the priority queue management in the depth-first kNN traversal introduces overhead that dominates when few points are retrieved. For large $k$ and full searches, however, the contiguous memory layout of the linear \Octree amortises this overhead through improved cache utilisation, explaining the crossover at $k \approx 300$--$500$. Our method is therefore best suited to applications that require large neighbourhoods or exhaustive cloud-wide queries, such as normal estimation with high $k$, density analysis, or batch processing pipelines. Beyond algorithmic efficiency and memory layout, the scalability of these methods in modern multi-core systems is also a crucial factor for large-scale point cloud processing, which is discussed in the following section.

\subsection{Parallel implementation and efficiency} % 7.4
We use OpenMP to parallelise the queries. In fixed-radius searches, we provide a \textit{kernel} and a radius that are independent of the chosen centre, and then we perform simple loop-level parallelisation over all the centres specified in $v_s$. For kNN searches, we essentially do the same for a given $k$, instead of radius and kernel. We distribute our iterations using OpenMP \textit{dynamic scheduling}, which allocates chunks of iterations to threads at runtime, with automatic block size, which adapts well to the unbalanced workloads inherent in spatial queries where different centres yield distinct numbers of neighbours. We found that it offers better performance on our system than other OpenMP scheduling strategies.

To wrap up the search efficiency analysis, we perform a scalability experiment on the structures. During the executions showcased in this section, we run our programs with the memory interleave policy using \texttt{numactl --interleave=all}. This makes the assignment of threads to nodes less relevant and minimises the Non-Uniform Memory Access (NUMA) architecture's impact. Without page interleaving and small trees, affinity becomes important: whether the threads are executing on the node containing the tree. 

\begin{figure*}[htpb]
    \centering
    \begin{subfigure}{0.48\textwidth} % struct
        \centering
        \caption{\neighboursStruct, subset searches}
        \label{subfig:lille0_parallelism_struct_subset}
        \includegraphics[width=\linewidth]{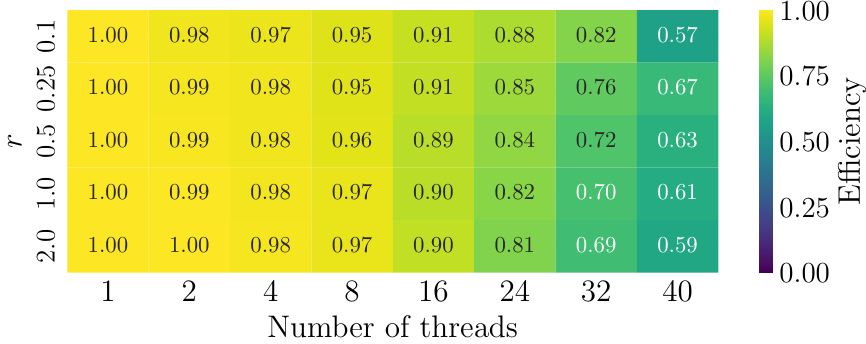}
        \vspace{2pt}
        \caption{\neighboursStruct, full searches}
        \label{subfig:lille0_parallelism_struct_full}
        \includegraphics[width=\linewidth]{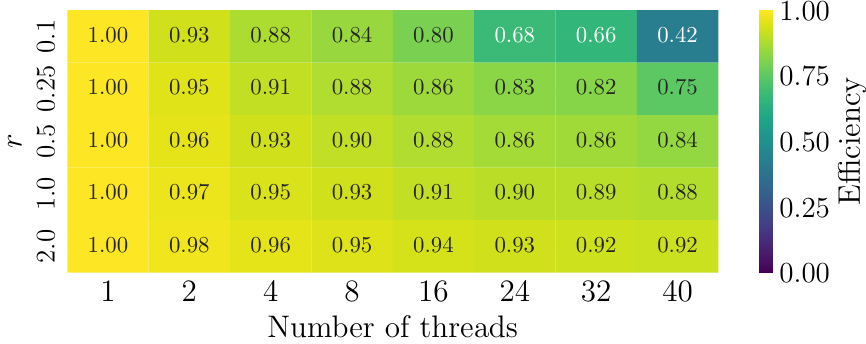}
    \end{subfigure}
    \hfill
    \begin{subfigure}{0.48\textwidth}  % nano
        \centering
        \caption{\neighboursNano, subset searches}
        \label{subfig:lille0_parallelism_nano_subset}
        \includegraphics[width=\linewidth]{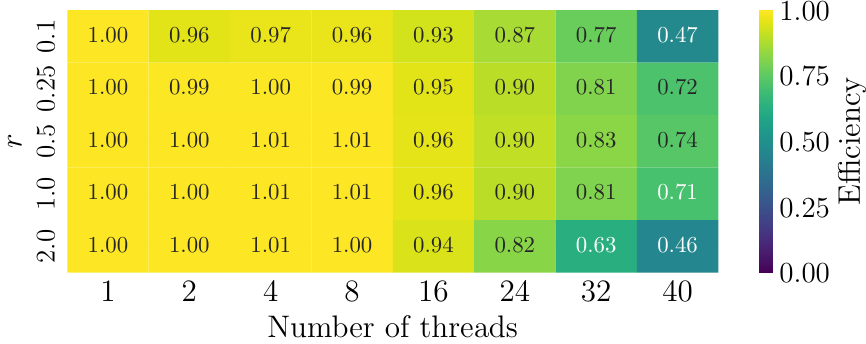}
        \vspace{2pt}
        \caption{\neighboursNano, full searches}
        \label{subfig:lille0_parallelism_nano_full}
        \includegraphics[width=\linewidth]{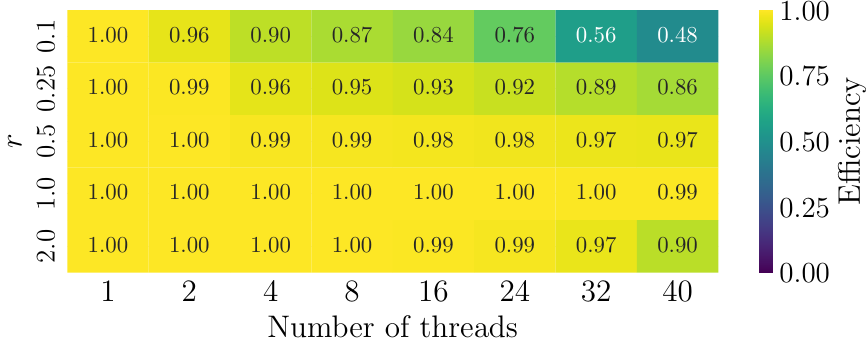}
    \end{subfigure}

    \caption{Parallel efficiency matrices for various cases on cloud \textit{Lille\_0}.}
    \label{fig:parallelism-lille-0}
\end{figure*}

The heatmaps of Figure~\ref{fig:parallelism-lille-0}, show how the parallel efficiency varies on cloud \textit{Lille\_0}, after Hilbert reordering. Executions with higher radii present better parallel efficiency, which may be because other threads find more points and better populate the shared caches of each NUMA package. Nanoflann achieves a similar parallel efficiency to our linear \Octree. 

Our parallelisation achieves almost ideal efficiency for full searches, as assigning consecutive centres to each thread greatly improves memory locality. Given a centre $p_{i}$ and a small $j$, centre $p_{i+j}$ is typically closer in space thanks to the reordering, and the intersection of their neighbourhood sets $\mathcal{N}_i \cap \mathcal{N}_{i+j}$, will typically be larger, making it more likely that points in $\mathcal{N}_{i+j}$ are already in cache from previous queries.

\subsection{Memory footprint and build times}~\label{sec:memory-footprint-and-build-times} % 7.5
In this section, we analyse the memory requirements for constructing some spatial search structures, and the time required to build them. In particular, we compare the baseline pointer \Octree (\ptrOctree), the optimised linear \Octree (\linOctree), unibnOctree (\unibnOctree), nanoflann's \KDTree (\nanoflannKDTree) and PCL's \KDTree (\pclKDTree). We do not include PCL's \Octree in this discussion because its resolution-based stopping criteria differ from the $N_{max}$ approach found in all the other structures.

We measured memory overhead across different clouds in two ways. First, by directly adding up the allocated sizes of all arrays and variables, and second, by using \textit{heaptrack}~\cite{heaptrack} to measure it. Table~\ref{tab:memory-measured} shows the results.

\begin{table*}[htb]
\centering
\caption{Measured memory consumption (MB). Columns are sorted by average relative overhead (efficiency). \pclKDTree{} could only be measured with \textit{heaptrack} as the library does not expose a simple API for retrieving memory usage.}\label{tab:memory-measured}
\footnotesize
%\scriptsize
\begin{tabular}{lllllll}
\toprule
Cloud (memory size) & \linOctree & \picotreeKDTree & \unibnOctree & \nanoflannKDTree & \ptrOctree & \pclKDTree \\
\midrule

\textit{5080\_54400} (372) 
& \textbf{26} & 61 & 91 & 110 & 156 & \textit{608} \\

\textit{bildstein\_station1} (906) 
& \textbf{68} & 148 & 217 & 267 & 394 & \textit{1477} \\

\textit{ParLux} (1525) 
& \textbf{99} & 246 & 355 & 448 & 625 & \textit{3800} \\

\textit{Lille} (3662) 
& \textbf{229} & 589 & 839 & 1073 & 1480 & \textit{6000} \\

\textit{sg27} (13110) 
& \textbf{1273} & 2152 & 2889 & 3827 & 6413 & \textit{21300} \\

\textit{Speulderbos} (22027) 
& \textbf{1492} & 3581 & 5506 & 6505 & 9179 & \textit{35900} \\

\addlinespace
\midrule

\textbf{Average overhead} 
& \textbf{7.29 \%} 
& 16.27 \% 
& 23.61 \% 
& 29.41 \% 
& 42.90 \% 
& \textit{177.49 \%} \\

\bottomrule
\end{tabular}
\end{table*}

The memory overhead depends on the ratio of the number of tree nodes to the number of points inserted, which in turn depends on cloud geometry and relative densities. Thus, we have performed a more careful analysis of the \Octrees. Given a fixed $N_{max}$, let $T_o$ be the node size, $T_p$ be the additional memory needed per point inserted, and $\rho$ be the ratio between the number of nodes and points inserted for that $N_{max}$. 

On average, for $N_{max} = 128$ we have found a value of $\rho = 0.045$ across our datasets for complete \Octrees (\ptrOctree{} and \linOctree{}) and $\rho = 0.029$ for non-complete \Octrees (\unibnOctree{}). Considering $32$ bytes per point, we can approximate total memory usage by $N(32 + T_p + \rho T_o)$, and memory overhead by Equation~\eqref{eq:memory_overhead}. 

\begin{equation}\label{eq:memory_overhead}
    \omega = \frac{T_p + \rho T_o}{32}
\end{equation} 

The choice of $N_{max} = 128$ represents a commonly used trade-off between tree depth and leaf occupancy. Smaller values increase tree depth and $\rho$, raising memory overhead, while larger values reduce node count but may degrade query performance in sparse neighbourhoods where few points per leaf are expected. The relative performance ordering among the tested structures is not expected to change qualitatively for moderate variations around this value, as the dominant factors are SFC ordering and memory layout.

In Table~\ref{tab:memory-expected} we show the values for each \Octree. Note that the expected overhead $\omega$ approximates well the measurements from Table~\ref{tab:memory-measured}.

\begin{table}[htb]
\centering
\caption{Expected overhead $\omega$ for \Octree structures over the evaluated datasets.}\label{tab:memory-expected}
\footnotesize
\begin{tabular}{lllll}
\toprule
Structure & $T_o$ & $T_p$ & $\rho$ & $\omega$ \\
\midrule
\ptrOctree 
& 120 & 8 & 0.045 & 41.87\% \\
\linOctree & 52 & 0 & 0.045 & \textbf{7.31\%} \\
\addlinespace
\unibnOctree
& 128 & 4 & 0.029 & 24.10\% \\
\bottomrule
\end{tabular}
\end{table}

From both benchmarks, it is clear that the linear \Octree offers the most compact representation. Our results show a $2\times$ improvement in compactness over \picotreeKDTree and a $3\times$ improvement over \unibnOctreeLib, the most memory-efficient \KDTree and pointer-based \Octree variants tested. Additionally, it benefits from being contained in a few arrays. In fact, \textit{heaptrack} shows that the linear \Octree requires only a few memory allocations, corresponding to the resizing of the \textit{leafs} array during parallel construction. Additionally, the other structures require allocations for each new node subdivision, and can be in the order of millions.

We can further compress the memory used by the linear \Octree structure if we do not precompute octant centres during initialisation, reducing $T_o$ to $28$ bytes. Note that finding an octant centre is relatively fast, as it only entails SFC decoding and a few floating-point operations.

The contiguous array layout of the linear \Octree also confers a structural advantage on NUMA systems. Pointer-based trees allocate each node independently, so nodes belonging to the same subtree may reside on different NUMA domains, incurring costly cross-domain memory traffic during traversal. In contrast, the linear \Octree is backed by a small number of contiguous arrays that are amenable to first-touch allocation policies, allowing the entire structure to be pinned within a single NUMA domain and exploiting hardware prefetchers more effectively.

\begin{table*}[htpb]
\centering
\caption{Time (s) to build each structure over different clouds after Hilbert reordering.}
\label{tab:build-times}
\footnotesize
\begin{tabular}{lccccccc}
\toprule
Cloud & 
\begin{tabular}[c]{@{}c@{}}\linOctree \\[-0.9ex] \scriptsize (parallel)\end{tabular} & 
\begin{tabular}[c]{@{}c@{}}\linOctree \\[-0.9ex] \scriptsize (sequential)\end{tabular} & 
\picotreeKDTree & 
\unibnOctree & 
\nanoflannKDTree & 
\ptrOctree & 
\pclKDTree \\
\midrule
\textit{5080\_54400} & \textbf{0.06} & 0.23 & 1.44 & 0.77 & 2.25 & 2.61 & 3.81 \\
\textit{bildstein\_station1} & \textbf{0.20} & 1.06 & 3.89 & 2.86 & 7.95 & 8.47 & 9.71 \\
\textit{ParLux} & \textbf{0.27} & 1.30 & 6.25 & 4.26 & 10.45 & 10.88 & 14.08 \\
\textit{Lille} & \textbf{0.65} & 2.91 & 18.36 & 11.58 & 31.38 & 30.30 & 37.81 \\
\textit{sg27} & \textbf{3.85} & 16.94 & 119.27 & 45.18 & 155.07 & 129.11 & 178.69 \\
\textit{Speulderbos} & \textbf{5.82} & 25.57 & 118.67 & 62.04 & 212.75 & 171.02 & 303.19 \\
\bottomrule
\end{tabular}
\end{table*}

Apart from being more compact, the linear \Octree is also faster to build, and thus suitable for tasks involving large clouds where latency is a concern. Table~\ref{tab:build-times} summarises our results. For comparison purposes, we tested the linear \Octree with both parallel and sequential construction. All other structures do not offer parallel construction.

From these measurements, we observe a consistent speedup of $2.5-4\times$ from the second-fastest \Octree, \unibnOctree, to the linear \Octree (sequentially built) and of $4-9\times$ from \picotreeKDTree, the fastest \KDTree. Parallelisation with 40 threads provides an additional improvement, with a measured average speedup of $4.53\times$.

\section{Conclusions}\label{sec:conclusions} % 8

Throughout this study, we present two primary improvements over traditional pointer-based \Octree structures for neighbourhood searching in 3D point clouds: (1) a reordering process based on three-dimensional Space-Filling Curves (SFCs), and (2) a linear \Octree structure with optimised query algorithms. Our results demonstrate that both approaches consistently reduce execution times for fixed-radius and kNN searches across several well-known datasets with diverse characteristics. Thus, these proposals offer a robust solution for enhancing the performance of large-scale 3D point cloud processing workflows. The main contributions of this work are:

\begin{itemize}
\item SFC reordering using Morton and Hilbert curves enhances memory locality and consistently achieves speedups of up to $50\%$ across the evaluated algorithms and data structures. Furthermore, point reordering significantly improves parallel scalability, enabling our implementation to reach an efficiency of up to $90\%$ on a 40-core system.
\item We introduce kNN locality histograms as a diagnostic metric for spatial locality, whose skewness correlates with cache miss rates and can be approximated on a small point subset to predict the best SFC ordering prior to index construction.
\item Our linear \Octree, combined with the \neighboursStruct method, delivers substantial performance gains compared to both pointer-based implementations and state-of-the-art libraries. These improvements are particularly significant in high-density scenarios ($\mu > 10^3$) and for large $k$ values ($k > 300$).
\item The proposed linear \Octree offers the most compact memory representation and the fastest construction times among the tested structures, making it the most suitable choice for large-scale or resource-constrained environments.
\end{itemize}

Adapting our methods for GPU acceleration is an identified future direction for our research, as the contiguous memory layout and the bulk parallelism of fixed-radius queries may be well suited to GPU architecture. However, this development presents several challenges to achieving optimal performance, such as managing irregular memory access patterns and thread divergence during hierarchical tree traversal.

\section*{Acknowledgements} % 9
The authors acknowledge CESGA (Centro de Supercomputación de Galicia) for providing access to the Finisterrae-III supercomputer, which enabled the computational resources necessary for the initial tests. This research was funded by the Agencia Estatal de Investigación (Spain) (MCIN/AEI/10.13039/501100011033) code PID2022-141623NB-I00, the Xunta de Galicia - Consellería de Cultura, Educación, Formación Profesional e Universidades (Centro de investigación de Galicia accreditation 2024-2027 ED431G-2023/04 and Reference Competitive Group accreditation ED431C-2022/016), the European Union (European Regional Development Fund - ERDF/EU).

\section*{CRediT authorship contribution statement}
\textbf{Pablo D. Viñambres:} Software, Formal analysis, Visualisation, Data Curation, Formal Analysis, Writing -- original draft. Writing -- review and editing.
\textbf{Miguel Yermo:} Conceptualisation, Methodology, Software, Supervision, Writing -- original draft, Writing -- review and editing.
\textbf{Silvia R. Alcaraz:} Conceptualisation, Writing -- review and editing.
\textbf{Oscar G. Lorenzo:} Conceptualisation, Writing -- review and editing.
\textbf{Francisco F. Rivera:} Conceptualisation, Supervision, Writing -- review and editing, Funding acquisition.
\textbf{José C. Cabaleiro:} Conceptualisation, Supervision, Writing -- review and editing, Funding acquisition.

\section*{Data availability}
The datasets and source code used in this study are publicly available. Sources and references are provided in the manuscript.

\bibliographystyle{elsarticle-num}
\bibliography{biblio}

\end{document}